\documentclass{article}

\usepackage{times}
\usepackage{graphicx} % more modern
\usepackage{subfigure}

\usepackage{natbib}

\usepackage{algorithm}
\usepackage{algorithmic}

\usepackage{hyperref}

\usepackage[accepted]{icml2016}

\usepackage{times}
\usepackage{epsfig}
\usepackage{graphicx}
\usepackage{amsmath}
\usepackage{amssymb}
\usepackage{algorithm}
\usepackage{algorithmic}
\usepackage{multirow}

\def\B{{\bf B}}
\def\b{{\bf b}}

\def\F{{\bf F}}

\def\G{{\bf G}}

\def\X{{\bf X}}
\def\Y{{\bf Y}}

\def\S{{\bf S}}

\def\x{{\bf x}}
\def\y{{\bf y}}

\def\V{{\bf V}}

\def\W{{\bf W}}
\def\w{{\bf w}}
\def\0{{\bf 0}}
\def\1{{\bf 1}}

\def\JM{{\mathcal J}}

\def\RB{{\mathbb R}}

\def\tr{\mathrm{tr}}

\def\sgn{\mathrm{sign}}

\icmltitlerunning{DCMH}

\begin{document}

\twocolumn[
\icmltitle{Deep Cross-Modal Hashing}

\icmlauthor{Qing-Yuan Jiang}{jiangqy@lamda.nju.edu.cn}
%\icmladdress{National Key Laboratory for Novel Software Technology,
%            Department of Computer Science and Technology, Nanjing University, Nanjing, China}
\icmlauthor{Wu-Jun Li}{liwujun@nju.edu.cn}
\icmladdress{National Key Laboratory for Novel Software Technology,\\
            Department of Computer Science and Technology, \\Nanjing University, Nanjing, China}

\icmlkeywords{learning to hash, multimodal hash, cross-modal retrieval}

\vskip 0.3in
]

\begin{abstract}
Due to its low storage cost and fast query speed, cross-modal hashing~(CMH) has been widely used for similarity search in multimedia retrieval applications. However, almost all existing CMH methods are based on hand-crafted features which might not be optimally compatible with the hash-code learning procedure. As a result, existing CMH methods with hand-crafted features may not achieve satisfactory performance. In this paper, we propose a novel cross-modal hashing method, called \underline{d}eep \underline{c}ross-\underline{m}odal \underline{h}ashing~(DCMH), by integrating feature learning and hash-code learning into the same framework. DCMH is an end-to-end learning framework with deep neural networks, one for each modality, to perform feature learning from scratch. Experiments on two real datasets with text-image modalities show that DCMH can outperform other baselines to achieve the state-of-the-art performance in cross-modal retrieval applications.
\end{abstract}

\section{Introduction}

% from ANN to Cross-Modal Hashing
Approximate nearest neighbor~(ANN) search~\cite{DBLP:journals/cacm/AndoniI08,DBLP:conf/stoc/AndoniR15} plays a fundamental role in machine learning and related applications like information retrieval. Due to its low storage cost and fast retrieval speed, hashing has recently attracted much attention from the ANN research community~\cite{DBLP:conf/nips/WeissTF08,DBLP:conf/nips/RaginskyL09,DBLP:conf/icml/WangKC10,DBLP:conf/icml/LiuWKC11,DBLP:conf/icml/NorouziF11,DBLP:conf/nips/0002FS12,DBLP:conf/icml/RastegariCFHD13,DBLP:conf/icml/YuKGC14,DBLP:conf/nips/LiuMKC14,DBLP:conf/icml/Shrivastava014,DBLP:conf/nips/AndoniILRS15,DBLP:conf/icml/NeyshaburS15,DBLP:conf/icml/LengWCZL15}. The goal of hashing is to map the data points from the original space into a Hamming space of binary codes where the similarity in the original space is preserved in the Hamming space. By using binary hash codes to represent the original data, the storage cost can be dramatically reduced. Furthermore, we can achieve a constant or sub-linear time complexity for search by using hash codes to construct an index. Hence, hashing has become more and more popular for ANN search in large-scale datasets.

In many applications, the data can have multi-modalities. For example, besides the image content, there also exists text information like tags for the images in Flickr and many other social websites. This kind of data is always called multi-modal data. With the rapid growth of multi-modal data in real applications especially multimedia applications, multi-modal hashing~(MMH) has recently been widely used for ANN search~(retrieval) on multi-modal datasets.

% Cross-Modal Hashing % Existing work and deep hashing
Existing MMH methods can be roughly divided into two main categories: \emph{mutli-source hashing}~(MSH)
~\cite{DBLP:conf/mm/SongYHSH11,DBLP:conf/sigir/ZhangWS11}
and \emph{cross-modal hashing}~(CMH)
~\cite{DBLP:conf/ijcai/KumarU11,DBLP:conf/cvpr/DingGZ14,DBLP:conf/aaai/ZhangL14,DBLP:conf/cvpr/LinDH015}.
The goal of MSH is to learn hash codes by utilizing all the information from multiple modalities. Hence, MSH requires that
all the modalities should be observed for all data points including the query points and those in database. In practice, the application of MSH is limited because in many cases it is difficult to acquire all the modalities of all data points. On the contrary, the application scenarios of CMH are more flexible than those of MSH. In CMH, the modality of a query point is different from the modality of the points in the database. Furthermore, typically the query point has only one modality and the points in the database can have one or more modalities.  For example, we can use text queries to retrieve images in the database, and we can also use image queries to retrieve texts in the database. Due to its wide application, CMH has gained more attention than MSH.

Many CMH methods have recently been proposed. Representative methods include cross modality similarity sensitive hashing~(CMSSH)~\cite{DBLP:conf/cvpr/BronsteinBMP10}, cross view hashing~(CVH)~\cite{DBLP:conf/ijcai/KumarU11},
multi-modal latent binary embedding~(MLBE)~\cite{DBLP:conf/kdd/ZhenY12},
co-regularized hashing~(CRH)~\cite{DBLP:conf/nips/ZhenY12}, semantic correlation maximization~(SCM)~\cite{DBLP:conf/aaai/ZhangL14},
collective matrix factorization hashing~(CMFH)~\cite{DBLP:conf/cvpr/DingGZ14},
semantic topic multi-modal hashing~(STMH)~\cite{DBLP:conf/ijcai/WangGWH15} and
semantics preserving hashing~(SePH)~\cite{DBLP:conf/cvpr/LinDH015}. Almost all these existing CMH methods are based on hand-crafted features. One shortcoming of these hand-crafted feature based methods is that the feature extraction procedure is independent of the hash-code learning procedure, which means that the hand-crafted features might not be optimally compatible with the hash-code learning procedure. Hence, these existing CMH methods with hand-crafted features may not achieve satisfactory performance in real applications.

Recently, deep learning with neural networks~\cite{DBLP:conf/nips/CunBDHHHJ89,DBLP:conf/nips/KrizhevskySH12} has been widely used to perform feature learning from scratch with promising performance. There also exist some methods which adopt deep learning for uni-modal hashing~\cite{DBLP:conf/cvpr/ZhaoHWT15,DBLP:conf/cvpr/LiongLWMZ15}. However, to the best of our knowledge, there has not appeared any deep CMH methods which can perform simultaneous feature learning and hash-code learning in the same framework.

% Deep Cross-Modal Hashing
In this paper, we propose a novel CMH method, called \underline{d}eep \underline{c}ross-\underline{m}odal \underline{h}ashing~(DCMH), for cross-modal retrieval applications. The main contributions of DCMH are outlined as follows:

\begin{itemize}
    \item DCMH is an end-to-end learning framework with deep neural networks, one for each modality, to perform feature learning from scratch.
    \item To the best of our knowledge, DCMH is the first CMH method which integrates both feature learning and hash-code learning into the same deep learning framework.
    \item The hash-code learning problem is essentially a discrete optimization problem, which is difficult to learn. Hence, most existing CMH methods typically solve this problem by relaxing the original discrete learning problem into a continuous learning problem. This relaxation procedure may deteriorate the accuracy of the learned hash codes~\cite{DBLP:conf/nips/LiuMKC14}. Unlike these relaxation-based methods, DCMH directly learns the discrete hash codes without relaxation.
    \item Experiments on real datasets with text-image modalities show that DCMH can outperform other baselines to achieve
        the state-of-the-art performance in cross-modal retrieval applications.
\end{itemize}

The rest of this paper is organized  as follows. Section~\ref{sec:def} introduces the problem definition of this paper.
We present our DCMH method in Section~\ref{sec:framework}, including the model formulation and learning algorithm. Experiments are shown in Section~\ref{sec:exp}. At last, we conclude our work in Section~\ref{sec:conclusion}.

\section{Problem Definition}
\label{sec:def}
In this section, we introduce the notation and problem definition of this paper.
\subsection{Notation}
Boldface lowercase letters like $\w$ are used to denote vectors.
Boldface uppercase letters like $\W$ are used to denote matrices, and the $(i,j)$th element of $\W$ is denoted as $W_{ij}$. The $i$th row of $\W$ is denoted as $\W_{i*}$, and the $j$th column of $\W$ is denoted as $\W_{*j}$. $\W^T$ is the transpose of $\W$.
We use $\textbf{1}$ to denote a vector with all elements being 1.
$\tr(\cdot)$ and $||\cdot||_F$ denote the trace of a matrix and the Frobenius norm of a matrix, respectively.
$\sgn(\cdot)$ is an element-wise sign function defined as follows:
$$ \sgn(x)=\left\{
\begin{aligned}
1\quad& x\geq 0, \\
-1\quad& x<0.
\end{aligned}
\right.
$$
\subsection{Cross-Modal Hashing}
Although the method proposed in this paper can be easily adapted to cases with more than two modalities, we only focus on the case with two modalities here.

Assume that we have $n$ training entities (data points), each of which has two modalities of features. Without loss of generality, we use text-image datasets for illustration in this paper, which means that each training point has both text modality and image modality. We use $\X=\{\x_i\}_{i=1}^n$ to denote the image
modality, where $\x_i$ can be the hand-crafted features or the raw pixels of image $i$.
Moreover, we use $\Y=\{\y_i\}_{i=1}^n$ to denote the text modality, where $\y_i$ is typically the tag information
related to image $i$. In addition, we are also given a cross-modal similarity matrix $\S$. $S_{ij}=1$ if image $\x_i$ and text $\y_j$ are similar, and $S_{ij}=0$ otherwise. Here, the similarity is typically defined by some semantic information such as class labels. For example, we can say that image $\x_i$ and text $\y_j$ are similar if they share the same class label. Otherwise, image $\x_i$ and text $\y_j$ are dissimilar if they are from different classes.

Given the above training information $\X$, $\Y$ and $\S$, the goal of cross-modal hashing is to learn two hash functions for the two modalities:
$h^{(x)}(\x) \in \{-1,+1\}^{c}$ for the image modality and $h^{(y)}(\y) \in \{-1,+1\}^{c}$ for the text modality, where $c$ is the length of binary code. These two hash functions should preserve the \emph{cross-modal similarity} in $\S$. More specifically, if $S_{ij}=1$, the Hamming distance between the binary codes $\b^{(x)}_i=h^{(x)}(\x_i) $ and $\b^{(y)}_j=h^{(y)}(\y_j) $
should be small. Otherwise if $S_{ij}=0$, the corresponding Hamming distance should be large.

\begin{figure*}[t]
\centering
\includegraphics[width=0.95\textwidth]{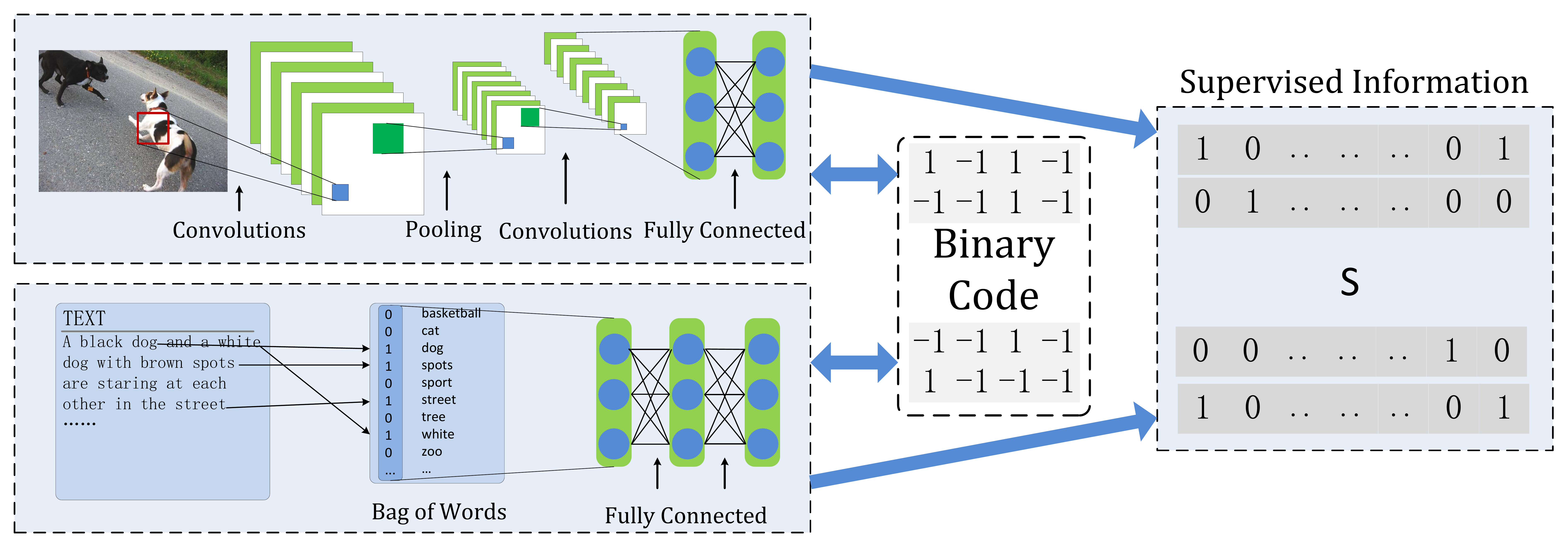}
\vspace{-0.2cm}
\caption{The end-to-end deep learning framework of our DCMH model.}
\label{fig:framework}
\vspace{-0.5cm}
\end{figure*}

Here, we assume that both modalities of features for each point in the \emph{training set} are observed although our method can also be easily adapted to other settings where some \emph{training points} have only one modality of features being observed. Please note that we only make this assumption for training points. After we have trained the model, we can use the learned model to generate hash codes for query and database points of either one modality or two modalities, which exactly matches the setting of cross-modal retrieval applications.

\section{Deep Cross-Modal Hashing}
\label{sec:framework}
In this section, we present the details about our deep CMH~(DCMH) method, including model formulation and learning algorithm.

\subsection{Model}

The whole DCMH model is shown in Figure~\ref{fig:framework}, which is an end-to-end learning framework by seamlessly integrating two parts: the feature learning part and the hash-code learning part. During learning, each part can give feedback to the other part.

\subsubsection{Feature Learning Part with Deep Neural Networks}
The feature learning part contains two deep neural networks, one for image modality and the other for text modality.

The deep neural network for image modality is a CNN model adapted from~\cite{DBLP:conf/bmvc/ChatfieldSVZ14}.
There are eight layers in this CNN model. The first seven layers are the same as those in CNN-F of~\cite{DBLP:conf/bmvc/ChatfieldSVZ14}. The eightth layer is a fully-connected layer with the output being the learned image features.

\begin{table}[t]
\small
\centering
\caption{Configuration of the CNN for image modality.}
\label{table:conf_img}
\begin{tabular}{|c|c|}
\hline
Layer&Configuration\\
\hline
\hline
conv1&f. $64\times 11\times 11$; st. $4\times 4$, pad $0$, LRN,$\times 2$ pool\\
\hline
conv2&f. $265\times 5\times 5$; st. $1\times 1$, pad $2$, LRN,$\times 2$ pool\\
\hline
conv3&f. $265\times 3\times 3$; st. $1\times 1$, pad $1$\\
\hline
conv4&f. $265\times 3\times 3$; st. $1\times 1$, pad $1$\\
\hline
conv5&f. $265\times 3\times 3$; st. $1\times 1$, pad $1$,$\times 2$ pool\\
\hline
full6&4096\\
\hline
full7&4096\\
\hline
full8&Hash code length $c$\\
\hline
\end{tabular}
\vspace{-0.5cm}
\end{table}

Table~\ref{table:conf_img} shows the detailed configuration of the CNN for image modality. More specifically, eight layers are divided into five convolutional layers and three fully-connected layers, which are denoted as ``conv1~-~conv5" and \mbox{``full6 - full8"} in Table~\ref{table:conf_img}, respectively.
Each convolutional layer is described by several aspects:
\begin{itemize}
\item ``f. $num\times size \times size$" denotes the number of convolution filters and their receptive field size.
\item ``st" denotes the convolution stride.
\item ``pad" denotes the number of pixels to add to each size of the input;
\item ``LRN" denotes whether Local Response Normalization (LRN)~\cite{DBLP:conf/nips/KrizhevskySH12} is applied or not.
\item ``pool" denotes the down-sampling factor.
\item The number in the fully connected layers, such as ``4096", denotes the number of nodes in that layer. It is also the dimensionality of the output at that layer.
\end{itemize}
All the first seven layers use the Rectified Linear Unit~(ReLU)~\cite{DBLP:conf/nips/KrizhevskySH12} as activation function. For the eighth layer, we choose identity function as the activation function.

To perform feature learning from text, we first represent each text $\y_j$ as a vector with bag-of-words~(BOW) representation. And then the bag-of-words vectors are used as the input to a deep neural network with three fully-connected layers, denoted as ``full1 - full3". The detailed configuration of the deep neural network for text is shown in Table~\ref{table:conf_txt}, where the configuration shows the number of nodes in each layer. The activation function for the first two layers is ReLU, and that for the third layer is the identity function.

\begin{table}[t]
\small
\centering
\caption{Configuration of the deep neural network for text modality.}
\label{table:conf_txt}
\begin{tabular}{|c|c|}
\hline
Layer&Configuration\\
\hline
\hline
full1&Length of BOW vector\\
\hline
full2&4096\\
\hline
full3&Hash code length $c$\\
\hline
\end{tabular}
\vspace{-0.5cm}
\end{table}

Please note that the main goal of this paper is to show that it is possible to design an end-to-end learning framework for cross-modal hashing by using deep neural networks for feature learning from scratch. But how to design different neural networks is not the focus of this paper. Other deep neural networks might also be used to perform feature learning for our DCMH model, which will be leaved for future study.

\subsubsection{Hash-Code Learning Part}
Let $f(\x_i;\theta_x)\in \RB^c$ denote the learned image feature for point $i$, which corresponds to the output of the CNN for image modality. Furthermore, let $g(\y_j;\theta_y)\in \RB^c$ denote the learned text feature for point $j$, which corresponds to the output of the deep neural network for text modality. Here, $\theta_x$ is the network parameter of the CNN for image modality, and $\theta_y$ is the network parameter of the deep neural network for text modality.

The objective function of DCMH is defined as follows:
\vspace{-0.2cm}
\begin{align}
\label{eq:obj}
\min_{\B,\B^{(x)},\B^{(y)},\theta_x,\theta_y}\;&\JM=-\sum_{i,j=1}^n(S_{ij}\Theta_{ij}-\log(1+e^{\Theta_{ij}})) \nonumber \\
    &+\gamma(||\B^{(x)}-\F||^2_F+||\B^{(y)}-\G||^2_F)\nonumber\\
    &+\eta(||\F\1||^2_F+||\G\1||^2_F) \\
    s.t.\hspace{0.3cm} &\B^{(x)}\in\{-1,+1\}^{c\times n},\nonumber \\
      &\B^{(y)}\in\{-1,+1\}^{c\times n},\nonumber \\
      &\B\in\{-1,+1\}^{c\times n},\nonumber \\
     & \B = \B^{(x)} = \B^{(y)},\nonumber
\end{align}
where $\F\in \RB^{c\times n}$ with $\F_{*i} = f(\x_i;\theta_x)$, $\G\in \RB^{c\times n}$ with $\G_{*j} = g(\y_j;\theta_y)$, $\Theta_{ij}=\frac{1}{2}\F_{*i}^T\G_{*j}$, $\B^{(x)}_{*i}$ is the binary hash code for image $\x_i$, $\B^{(y)}_{*j}$ is the binary hash code for text $\y_j$, $\gamma$ and $\eta$ are hyper-parameters.

The first term $-\sum_{i,j=1}^n(S_{ij}\Theta_{ij}-\log(1+e^{\Theta_{ij}}))$ in~(\ref{eq:obj}) is the negative log likelihood of the cross-modal similarities with the likelihood function defined as follows:
$$ p(S_{ij}|\F_{*i},\G_{*j})=\left\{
\begin{aligned}
&\sigma(\Theta_{ij})  & S_{ij}=1 \\
&1-\sigma(\Theta_{ij})   & S_{ij}=0 \\
\end{aligned}
\right.
$$
where $\Theta_{ij}=\frac{1}{2}\F_{*i}^T\G_{*j}$ and $\sigma(\Theta_{ij})=\frac{1}{1+e^{-\Theta_{ij}}}$.

It is easy to find that minimizing this negative log likelihood, which is equivalent to maximizing the likelihood, can make the similarity~(inner product) between $\F_{*i}$ and $\G_{*j}$ be large when $S_{ij} = 1$ and be small when $S_{ij} = 0$. Hence, optimizing the first term in~(\ref{eq:obj}) can preserve the cross-modal similarity in $\S$ with the image feature representation $\F$ and text feature representation $\G$.

By optimizing the second term $\gamma(||\B^{(x)}-\F||^2_F+||\B^{(y)}-\G||^2_F)$ in~(\ref{eq:obj}), we can get $\B^{(x)}= \sgn(\F)$ and $\B^{(y)}= \sgn(\G)$. Hence, we can consider $\F$ and $\G$ to be the continuous surrogate of $\B^{(x)}$ and $\B^{(y)}$, respectively. Because $\F$ and $\G$ can preserve the cross-modal similarity in $\S$, the binary hash codes $\B^{(x)}$ and $\B^{(y)}$ can also be expected to preserve the cross-modal similarity in $\S$, which exactly matches the goal of cross-modal hashing.

The third term $\eta(||\F\1||^2_F+||\G\1||^2_F)$ in~(\ref{eq:obj}) is used to make each bit of the hash code be balanced on all the training points. More specifically, the number of $+1$ and that of $-1$ for each bit on all the training points should be almost the same. This constraint can be used to maximize the information provided by each bit.

In our experiment, we find that better performance can be achieved if the binary codes from the two modalities are set to be the same for the training points. Hence, we add another constraint $\B = \B^{(x)} = \B^{(y)}$ in~(\ref{eq:obj}). With this constraint, the problem in~(\ref{eq:obj}) can be equivalently transformed to the following reduced formulation:
\vspace{-0.2cm}
\begin{align}
\label{eq:obj2}
\min_{\B,\theta_x,\theta_y}\;\JM&=-\sum_{i,j=1}^n(S_{ij}\Theta_{ij}-\log(1+e^{\Theta_{ij}})) \nonumber \\
    &+\gamma(||\B-\F||^2_F+||\B-\G||^2_F)\nonumber\\
    &+\eta(||\F\1||^2_F+||\G\1||^2_F) \\
    s.t.\hspace{0.3cm} \B&\in\{-1,+1\}^{c\times n},\nonumber \\
                       \F& = f(\X;\theta_x) , \nonumber \\
                       \G& = g(\Y;\theta_y), \nonumber
\end{align}
where $\F = f(\X;\theta_x)$ means $\F_{*i} = f(\x_i;\theta_x)$, $\G = g(\Y;\theta_y)$ means $\G_{*j} = g(\y_j;\theta_y)$. This is the final objective function of our DCMH for learning.

From~(\ref{eq:obj2}), we can find that the parameters of the deep neural networks~($\theta_x$ and $\theta_y$) and the binary hash code~($\B$) are learned from the same objective function. That is to say, DCMH integrates both feature learning and hash-code learning into the same deep learning framework.

Please note that we only make $\B^{(x)} = \B^{(y)}$ for the training points. After we have learned the problem in~(\ref{eq:obj2}), we still need to generate different binary codes $\b^{(x)}_i=h^{(x)}(\x_i) $ and $\b^{(y)}_i=h^{(y)}(\y_i)$ for the two different modalities of the same point $i$ if point $i$ is a query point or a point from the database rather than a training point. This will be further illustrated in Section~\ref{sec:outOfSample}.

\subsection{Learning}

We adopt an alternating learning strategy to learn $\theta_x$, $\theta_y$ and $\B$. Each time we optimize one parameter with
the other parameters fixed. The whole alternating learning algorithm for DCMH is briefly outlined in
Algorithm~\ref{alg:DCMH}, and the detailed derivation will be introduced in the following content of this subsection.

\subsubsection{Fix $\theta_y$ and $\B$, optimize $\theta_x$}

When $\theta_y$ and $\B$ are fixed, we learn the CNN parameter $\theta_x$ of the image modality by using a
back-propagation~(BP) algorithm. As most existing deep learning methods~\cite{DBLP:conf/nips/KrizhevskySH12}, we utilize stochastic gradient descent~(SGD) to learn $\theta_x$  with the BP algorithm. More specifically, in each iteration we sample a mini-batch of points from the training set and then carry out our learning algorithm based on the sampled data.

In particular, for each sampled point $\x_i$, we first compute the following gradient:
\vspace{-0.1cm}
\begin{align}
\label{for:deri1}
\frac{\partial \JM}{\partial\F_{*i}}=&\frac{1}{2}\sum_{j=1}^n(\sigma(\Theta_{ij})\G_{*j}-S_{ij}\G_{*j})\nonumber\\
                                    &+2\gamma(\F_{*i}-\B_{*i})+2\eta\F\1.
\end{align}

Then we can compute $\frac{\partial \JM}{\partial\theta_{x}}$ with $\frac{\partial \JM}{\partial\F_{*i}}$ by using the chain rule, based on which BP can be used to update the parameter $\theta_{x}$.

\subsubsection{Fix $\B$ and $\theta_x$, optimize $\theta_y$}
When $\B$ and $\theta_x$ are fixed, we also learn the neural network parameter $\theta_y$ of the text modality by using SGD with a
BP algorithm. More specifically, for each sampled point $\y_j$, we first compute the following gradient:
\vspace{-0.1cm}
\begin{align}
\label{for:deri2}
\frac{\partial \JM}{\partial\G_{*j}}=&\frac{1}{2}\sum_{i=1}^n(\sigma(\Theta_{ij})\F_{*i}-S_{ij}\F_{*i})\nonumber\\
                                    &+2\gamma(\G_{*j}-\B_{*j})+2\eta\G\1.
\end{align}

Then we can compute $\frac{\partial \JM}{\partial\theta_{y}}$ with $\frac{\partial \JM}{\partial\G_{*j}}$ by using the chain rule, based on which BP can be used to update the parameter $\theta_{y}$.

\subsubsection{Fix $\theta_x$ and $\theta_y$, optimize $\B$}

When $\theta_x$ and $\theta_y$ are fixed, the problem in~(\ref{eq:obj2}) can be reformulated as follows:
\vspace{-0.1cm}
\begin{align}
\max_{\B}\;&\tr(\B^T(\gamma(\F+\G)))=\tr(\B^T\V)=\sum_{i,j}B_{ij}V_{ij}\nonumber\\
s.t.\hspace{0.3cm} &\B\in\{-1,+1\}^{c\times n}, \nonumber
\end{align}
where $\V=\gamma(\F+\G)$.

It is easy to find that the binary code $B_{ij}$ should keep the same sign as $V_{ij}$.
Therefore, we have:
\vspace{-0.1cm}
\begin{align}
\label{sol1:B}
\B=\sgn(\V)=\sgn(\gamma(\F+\G)).
\end{align}

\begin{algorithm}[tb]
\small
\caption{The learning algorithm for DCMH.}
\label{alg:DCMH}
\begin{algorithmic}
\REQUIRE Image set $\X$, text set $\Y$, and cross-modal similarity matrix $\S$.
\ENSURE Parameters $\theta_x$ and $\theta_y$ of the deep neural networks, and binary code matrix $\B$.
\STATE \textbf{Initialization}\\Initialize neural network parameters $\theta_x$ and $\theta_y$,
mini-batch size $N_x=N_y=128$, and iteration number $t_x=n/N_x,t_y=n/N_y$.
\REPEAT
    \FOR {$iter=1,2,\cdots,t_x$}
        \STATE Randomly sample $N_x$ points from $\X$ to construct a mini-batch.
        \STATE For each sampled point $\x_i$ in mini-batch, calculate $\F_{*i}=f(\x_i;\theta_x)$ by forward propagation.
        \STATE Calculate the derivative according to~(\ref{for:deri1}).
        \STATE Update the parameter $\theta_x$ by using back propagation.
    \ENDFOR
    \FOR {$iter=1,2,\cdots,t_y$}
        \STATE Randomly sample $N_y$ points from $\Y$ to construct a mini-batch.
        \STATE For each sampled point $\y_j$ in mini-batch, calculate $\G_{*j}=g(\y_j;\theta_y)$ by forward propagation.
        \STATE Calculate the derivative according to~(\ref{for:deri2}).
        \STATE Update the parameter $\theta_y$ by using back propagation.
    \ENDFOR
    \STATE Optimize $\B$ according to~(\ref{sol1:B}).
\UNTIL a fixed number of iterations
\end{algorithmic}
\end{algorithm}

\subsection{Out-of-Sample Extension} \label{sec:outOfSample}
For any point which is not in the training set, we can obtain its hash code as long as one of its modalities~(image or text) is observed. In particular, given the image modality $\x_q$ of point $q$, we can adopt forward propagation to generate the hash code as follows:
$$\b_q^{(x)}=h^{(x)}(\x_q)=\sgn(f(\x_q;\theta_x)).$$

Similarly, if point $q$ only has the text modality $\y_q$, we can also generate the hash code $\b_q^{(y)}$ as follows:
$$\b_q^{(y)}=h^{(y)}(\y_q)=\sgn(g(\y_q;\theta_y)).$$

Hence, our DCMH model can be used for cross-modal search where the query points have one modality and the points in database have the other modality.

%. leverage the function $h^{(2)}(\y_q)=\sgn(g(\y_q))$ to obtain the hash code $\b_q^{(2)}$ for text view.
%On the other hand, if we observe the two views for an unseen instance, we can use equation ~\ref{sol1:B} to obtain the hash code jointly.
%i.e., $\b_q=\sgn(\lambda f(\x_q)+\gamma g(\y_q))$.
\section{Experiment}
\label{sec:exp}

We carry out experiments on text-image datasets to verify the effectiveness of DCMH. DCMH is implemented with the open source deep learning toolbox MatConvNet~\cite{DBLP:conf/mm/VedaldiL15} on a NVIDIA K40 GPU server.

\subsection{Datasets}
Two datasets, \mbox{\emph{MIRFLICKR-25K}}~\cite{huiskes08} and {\emph{NUS-WIDE}}~\cite{DBLP:conf/civr/ChuaTHLLZ09}, are used for evaluation.

The original {\emph{MIRFLICKR-25K}} dataset~\cite{huiskes08} consists of 25,000 images collected from Flickr website.
Each image is associated with several textual tags. Hence, each point is a text-image pair. We select those points which have at least 20 textual tags for our experiment, and subsequently we get 20,015 points for our experiment. The text for each point is represented as a 1386-dimensional bag-of-words vector.
For the hand-crafted feature based method, each image is represented by a 512-dimensional SIFT feature vector.
Furthermore, each point is manually annotated with one of the 24 unique labels. The image $i$ and text $j$ are considered to be similar if point $i$ and point $j$ share the same label. Otherwise, they are considered to be dissimilar.

The {\emph{NUS-WIDE}} dataset~\cite{DBLP:conf/civr/ChuaTHLLZ09} contains 260,648 web images, and some images are associated with textual tags. It is a multi-label dataset where each point is annotated with one or multiple labels from 81 concept labels. We select 186,577 text-image pairs that belong to the 10 most frequent concepts. The text for each point is represented as a 1000-dimensional  bag-of-words vector. The hand-crafted feature for each image is a 500-dimensional bag-of-visual words~(BOVW) vector. The image $i$ and text $j$ are considered to be similar if point $i$ and point $j$ share at least one concept label. Otherwise, they are considered to be dissimilar.

\subsection{Evaluation Protocol and Baseline}

\subsubsection{Evaluation Protocol}

For {\emph{MIRFLICKR-25K}} dataset, we take 2000 data points as the test~(query) set
and the remaining points as the retrieval set~(database). For {\emph{NUS-WIDE}} dataset, we take 1\% of the dataset as the test~(query) set and the rest as the retrieval set. Moreover, we take 5000 data points from the retrieval set to construct the training set for both {\emph{MIRFLICKR-25K}} and {\emph{NUS-WIDE}}. The ground-truth neighbors are defined as those text-image pairs which share at least one semantic label.

For hashing-based retrieval, \emph{Hamming ranking} and \emph{hash lookup} are two widely used retrieval procedures~\cite{DBLP:conf/nips/LiuMKC14}. We also adopt these two procedures to evaluate our method and other baselines. The Hamming ranking procedure ranks the points in the database~(retrieval set) according to their Hamming distances to the given query point, in an increasing order. Mean average precision~(MAP)~\cite{DBLP:conf/nips/LiuMKC14} is the widely used metric to measure the accuracy of the Hamming ranking procedure. The hash lookup procedure returns all the points within a certain Hamming radius away from the query point. The precision-recall curve and F-measure~\cite{DBLP:conf/nips/LiuMKC14} are widely used metrics to measure the accuracy of the hash lookup procedure.

%In practice, Hamming Lookup can be achieved in constant time.
%We report the precision-recall and F-measure with different hamming radius $r,r\in\{0,1,\dots,c-1\}$ for Hamming Lookup task.
%We can use the formula~\ref{formula:pr} to calculate precision and recall,
%\begin{align}
%\label{formula:pr}
%    P_{r=i}&=\frac{\#\{\text{retrieved instances in hamming radius} r \}\bigcap\{\text{relevant instances}\}}{\#\text{retrieved instances in hamming radius} r}\nonumber\\
%    R_{r=i}&=\frac{\#\{\text{retrieved instances in hamming radius} r \}\bigcap\{\text{relevant instances}\}}{\#\text{relevant instances}}\\
%\end{align}
%Furthermore, we report the mAP for Hamming Ranking Task.
%mAP can be calculated using the following formula \ref{formula:map}:

%\begin{equation}
%\label{formula:map}
%    mAP=\frac{1}{n_q}\sum_{i=1}^{n_q} AP(\q_i)
%\end{equation}
%
%Where $\q_i$ denotes the $i$th query and $AP(\q_i)$ denotes the average precision (AP).
%Average precision is defined as $AP(\q)=\frac{1}{L}\sum_{i=1}^{n}P_q(r)\delta_q(r)$,
%where $L$ is the number of ground-truth neighbors for the query $\q$ among the retrieval items,
%$P_q(r)$ denotes the precision of the top $r$ retrieved entities, and $\delta(r)$ is a indicator function which
%equals 1 when the $r$th is a ground-truth neighbor and otherwise 0.

\subsubsection{Baseline}

Five state-of-the-art cross-modal hashing methods are adopted as baselines for comparison, including SePH~\cite{DBLP:conf/cvpr/LinDH015}, STMH~\cite{DBLP:conf/ijcai/WangGWH15}, SCM~\cite{DBLP:conf/aaai/ZhangL14}, CMFH~\cite{DBLP:conf/cvpr/DingGZ14} and CCA~\cite{hotelling1936relations}.
Source codes of SePH, STMH and SCM are kindly provided by the corresponding authors. While for CMFH and CCA whose codes are not available, we implement them carefully by ourselves. SePH is a kernel-based method, for which we use RBF kernel and take 500 randomly selected points as kernel bases by following its authors' suggestion. In SePH, the authors propose two strategies to construct the hash codes for retrieval~(database) points according to whether both modalities of a point are observed or not. However, in this paper we can only use one modality for the database~(retrieval) points, because the focus of this paper is on cross-modal retrieval. All the other parameters for all baselines are set according to the suggestion of the original papers of these baselines.

For our DCMH, we use a validation set to choose the hyper-parameter $\gamma$ and $\eta$, and find that good performance can be achieved with $\gamma=\eta=1$. Hence, we set $\gamma=\eta =1$ for all our experiments. We exploit the CNN-F network~\cite{DBLP:conf/bmvc/ChatfieldSVZ14} pre-trained on ImageNet dataset~\cite{DBLP:journals/corr/RussakovskyDSKSMHKKBBF14} to initialize the first seven layers of the CNN for image modality, and all the other parameters of the deep neural networks in DCMH are randomly initialized. The input for the image modality is the raw pixels, and that for the text modality is the BOW vectors. We fix the mini-batch size to be 128 and set the iteration number of the outer-loop in Algorithm~\ref{alg:DCMH} to be 500.

\subsection{Accuracy}
We report the accuracy for both Hamming ranking procedure and hash lookup procedure.
\subsubsection{Hamming Ranking}
The MAP results for DCMH and other baselines with hand-crafted features on \emph{MIRFLICKR-25K} and \emph{NUS-WIDE} are reported in Table~\ref{table:mir_map_ori} and Table~\ref{table:nus_map_ori}, respectively.
We can find that DCMH can outperform all the other baselines with hand-crafted features.

\begin{table}[h]
\small
\centering
\caption{Comparison to baselines with hand-crafted features on {{\emph{MIRFLICKR-25K}}} in terms of MAP. The best accuracy is shown in boldface.}
\label{table:mir_map_ori}
\begin{tabular}{|c|c|c|c|c|}
\hline
\multirow{2}{*}{Task} & \multirow{2}{*}{Method} & \multicolumn{3}{|c|}{code length} \\
\cline{3-5}
& &16 bits &32 bits & 64 bits\\
\hline
\hline
& DCMH & \textbf{0.7127} & \textbf{0.7203} & \textbf{0.7303}\\
\cline{2-5}
\multirow{3}{*}{Image Query}
& SePH & 0.6441 & 0.6492 & 0.6508\\
\cline{2-5}
\multirow{3}{*}{v.s.}
& STMH & 0.5876 & 0.5951 & 0.5942\\
\cline{2-5}
\multirow{3}{*}{Text Database}
& SCM & 0.6153 & 0.6279 & 0.6288\\
\cline{2-5}
& CMFH & 0.5804 & 0.5790 & 0.5797\\
\cline{2-5}
& CCA & 0.5634 & 0.5630 & 0.5626\\
\hline
& DCMH & \textbf{0.7504} & \textbf{0.7574} & \textbf{0.7704}\\
\cline{2-5}
\multirow{3}{*}{Text Query}
& SePH & 0.6455 & 0.6474 & 0.6506\\
\cline{2-5}
\multirow{3}{*}{v.s.}
& STMH & 0.5763 & 0.5877 & 0.5826\\
\cline{2-5}
\multirow{3}{*}{Image Database}
& SCM & 0.6102 & 0.6184 & 0.6192\\
\cline{2-5}
& CMFH & 0.5782 & 0.5778 & 0.5779\\
\cline{2-5}
& CCA & 0.5639 & 0.5631 & 0.5627\\
\hline
\end{tabular}
\vspace{-0.5cm}
\end{table}

\begin{table}[h]
\small
\centering
\caption{Comparison to baselines with hand-crafted features on {{\emph{NUS-WIDE}}} in terms of MAP. The best accuracy is shown in boldface.}
\label{table:nus_map_ori}
\begin{tabular}{|c|c|c|c|c|}
\hline
\multirow{2}{*}{Task} & \multirow{2}{*}{Method} & \multicolumn{3}{|c|}{code length}\\
\cline{3-5}
& &16 bits &32 bits & 64 bits\\
\hline
\hline
& DCMH & \textbf{0.6249} & \textbf{0.6355} & \textbf{0.6438}\\
\cline{2-5}
\multirow{3}{*}{Image Query}
& SePH & 0.5314 & 0.5340 & 0.5429\\
\cline{2-5}
\multirow{3}{*}{v.s.}
& STMH & 0.4344 & 0.4461 & 0.4534\\
\cline{2-5}
\multirow{3}{*}{Text Database}
& SCM & 0.4904 & 0.4945 & 0.4992\\
\cline{2-5}
& CMFH & 0.3825 & 0.3858 & 0.3890\\
\cline{2-5}
& CCA & 0.3742 & 0.3667 & 0.3617\\
\hline
& DCMH & \textbf{0.6791} & \textbf{0.6829} & \textbf{0.6906}\\
\cline{2-5}
\multirow{3}{*}{Text Query}
& SePH & 0.5086 & 0.5055 & 0.5170\\
\cline{2-5}
\multirow{3}{*}{v.s.}
& STMH & 0.3845 & 0.4089 & 0.4181\\
\cline{2-5}
\multirow{3}{*}{Image Database}
& SCM & 0.4595 & 0.4650 & 0.4691\\
\cline{2-5}
& CMFH & 0.3915 & 0.3944 & 0.3990\\
\cline{2-5}
& CCA & 0.3731 & 0.3661 & 0.3613\\
\hline
\end{tabular}
\end{table}

To further verify the effectiveness of DCMH, we exploit the CNN-F deep network~\cite{DBLP:conf/bmvc/ChatfieldSVZ14} pre-trained on ImageNet dataset, which is the same as the initial CNN of the image modality in DCMH, to extract CNN features. All the baselines are trained based on these CNN features. The MAP results for DCMH and other baselines with CNN features on \emph{MIRFLICKR-25K} and \emph{NUS-WIDE} are reported in Table~\ref{table:mir_map_vgg} and Table~\ref{table:nus_map_vgg}, respectively. We can find that DCMH can outperform all the other baselines except SePH. For SePH, DCMH can outperform it in most cases except the image to text retrieval on NUS-WIDE. Please note that SePH is a kernel-based method, which constructs kernels based on the CNN-F image features and text features. However, our DCMH can be seen as a linear method with deep features because the final layers of both modalities are fully-collected ones with identity activation functions. We find that the better performance of SePH mainly comes from the kernel features of SePH, which is verified by the worse results of a linear variant of SePH without kernels called ``SePH-linear" in Table~\ref{table:nus_map_vgg}. DCMH can outperform SePH with linear features in all cases. And even for SePH with kernel features, DCMH can outperform it for most cases. Hence, compared with these baselines with CNN-F features, the better accuracy of DCMH verifies that integrating both feature learning and hash-code learning into the same framework may improve the performance.

\begin{table}[h]
\small
\centering
\caption{Comparison to baselines with CNN-F features on {{\emph{MIRFLICKR-25K}}} in terms of MAP. The best accuracy is shown in boldface.}
\label{table:mir_map_vgg}
\begin{tabular}{|c|c|c|c|c|}
\hline
\multirow{2}{*}{Task} & \multirow{2}{*}{Method} & \multicolumn{3}{|c|}{code length} \\
\cline{3-5}
& &16 bits &32 bits & 64 bits\\
\hline
\hline
& DCMH & \textbf{0.7150} & \textbf{0.7203} & \textbf{0.7303}\\
\cline{2-5}
\multirow{3}{*}{Image Query}
& SePH & 0.7090 & 0.7110 & 0.7169\\
\cline{2-5}
\multirow{3}{*}{v.s.}
& STMH & 0.6242 & 0.6294 & 0.6314\\
\cline{2-5}
\multirow{3}{*}{Text Database}
& SCM & 0.6281 & 0.6286 & 0.6367\\
\cline{2-5}
& CMFH & 0.5761 & 0.5798 & 0.5807\\
\cline{2-5}
& CCA & 0.5619 & 0.5616 & 0.5616\\
\hline
& DCMH & \textbf{0.7545} & \textbf{0.7574} & \textbf{0.7704}\\
\cline{2-5}
\multirow{3}{*}{Text Query}
& SePH & 0.7127 & 0.7261 & 0.7309\\
\cline{2-5}
\multirow{3}{*}{v.s.}
& STMH & 0.6137 & 0.6196 & 0.6221\\
\cline{2-5}
\multirow{3}{*}{Image Database}
& SCM & 0.6068 & 0.6089 & 0.6108\\
\cline{2-5}
& CMFH & 0.5776 & 0.5792 & 0.5834\\
\cline{2-5}
& CCA & 0.5628 & 0.5630 & 0.5630\\
\hline
\end{tabular}
\vspace{-0.5cm}
\end{table}

\begin{table}[h]
\small
\centering
\caption{Comparison to baselines with CNN-F features on {{\emph{NUS-WIDE}}} in terms of MAP. The best accuracy is shown in boldface.}
\label{table:nus_map_vgg}
\begin{tabular}{|c|c|c|c|c|}
\hline
\multirow{2}{*}{Task} & \multirow{2}{*}{Method} & \multicolumn{3}{|c|}{code length}\\
\cline{3-5}
& &16 bits &32 bits & 64 bits\\
\hline
\hline
& DCMH & 0.6249 & 0.6355 & 0.6438\\
\cline{2-5}
\multirow{3}{*}{Image Query}
& SePH & \textbf{0.6538} & \textbf{0.6668} & \textbf{0.6720}\\
\cline{2-5}
\multirow{3}{*}{v.s.}
& SePH-linear & 0.6179 & 0.6306 & 0.6411 \\
\cline{2-5}
\multirow{3}{*}{Text Database}
& STMH & 0.5054 & 0.5277 & 0.5248\\
\cline{2-5}
& SCM & 0.4867 & 0.4958 & 0.4724\\
\cline{2-5}
& CMFH & 0.3997 & 0.3889 & 0.3972\\
\cline{2-5}
& CCA & 0.3783 & 0.3704 & 0.3649\\
\hline
& DCMH & \textbf{0.6791} & \textbf{0.6829} & \textbf{0.6906}\\
\cline{2-5}
\multirow{2}{*}{Text Query}
& SePH & 0.6600 & 0.6676 & 0.6854\\
\cline{2-5}
\multirow{3}{*}{v.s.}
& SePH-linear & 0.6369 & 0.6470 & 0.6551 \\
\cline{2-5}
\multirow{3}{*}{Image Database}
& STMH & 0.4717 & 0.5013 & 0.4993\\
\cline{2-5}
& SCM & 0.4280 & 0.4374 & 0.4090\\
\cline{2-5}
& CMFH & 0.3902 & 0.3870 & 0.3910\\
\cline{2-5}
& CCA & 0.3788 & 0.3705 & 0.3651\\
\hline
\end{tabular}
\vspace{-0.2cm}
\end{table}

\begin{figure*}[t]
\centering
\small
\begin{tabular}{c@{ }@{ }c@{ }@{ }c@{ }@{ }c}
\begin{minipage}{0.23\linewidth}\centering
    \includegraphics[width=1\textwidth]{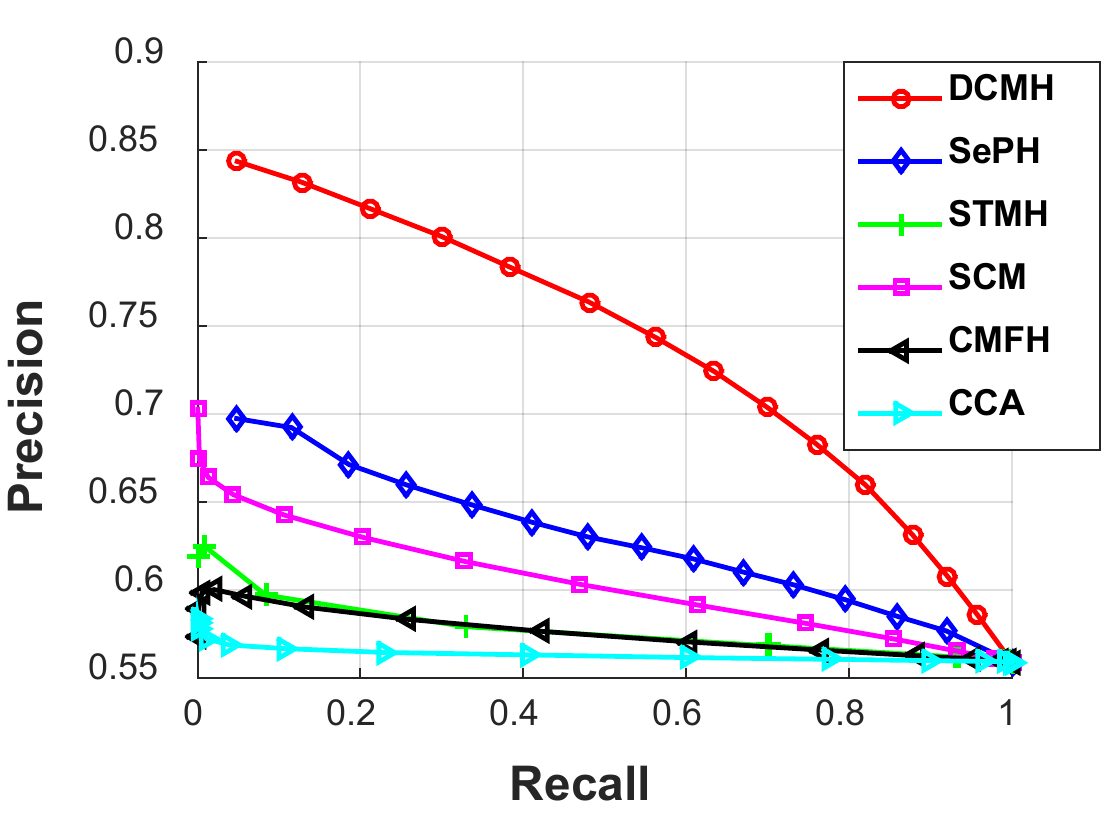}\\
    (a) Image $\to$ Text @MIRFLICKR-25K
\end{minipage} &
\begin{minipage}{0.23\linewidth}\centering
    \includegraphics[width=1\textwidth]{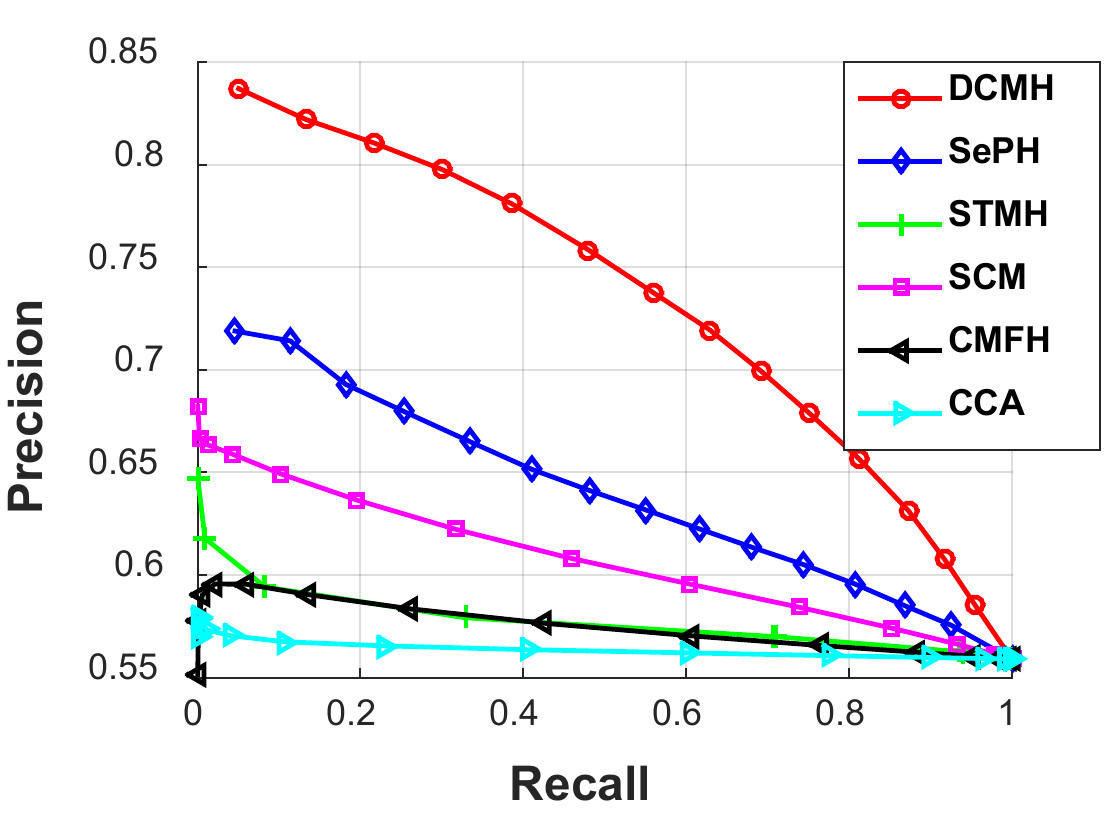}\\
    (b) Text $\to$ Image @MIRFLICKR-25K
\end{minipage} &
\begin{minipage}{0.23\linewidth}\centering
    \includegraphics[width=1\textwidth]{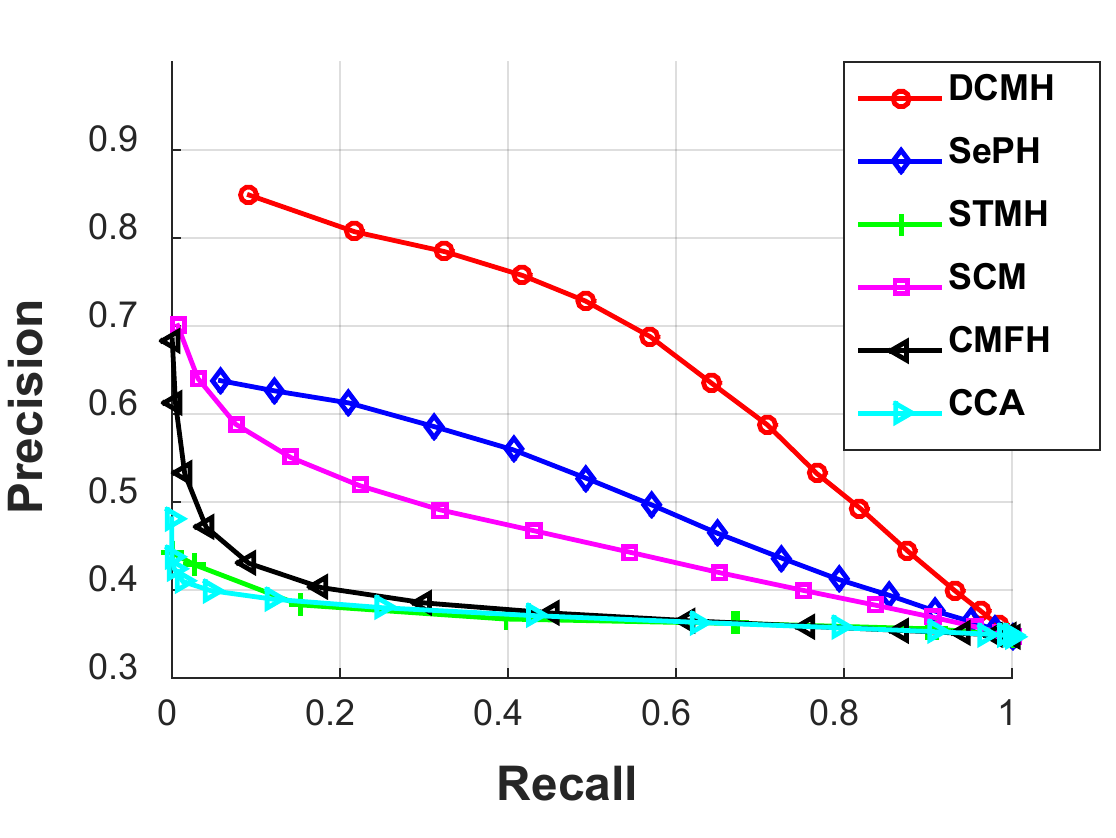}\\
    (c) Image $\to$ Text @NUS-WIDE
\end{minipage} &
\begin{minipage}{0.23\linewidth}\centering
    \includegraphics[width=1\textwidth]{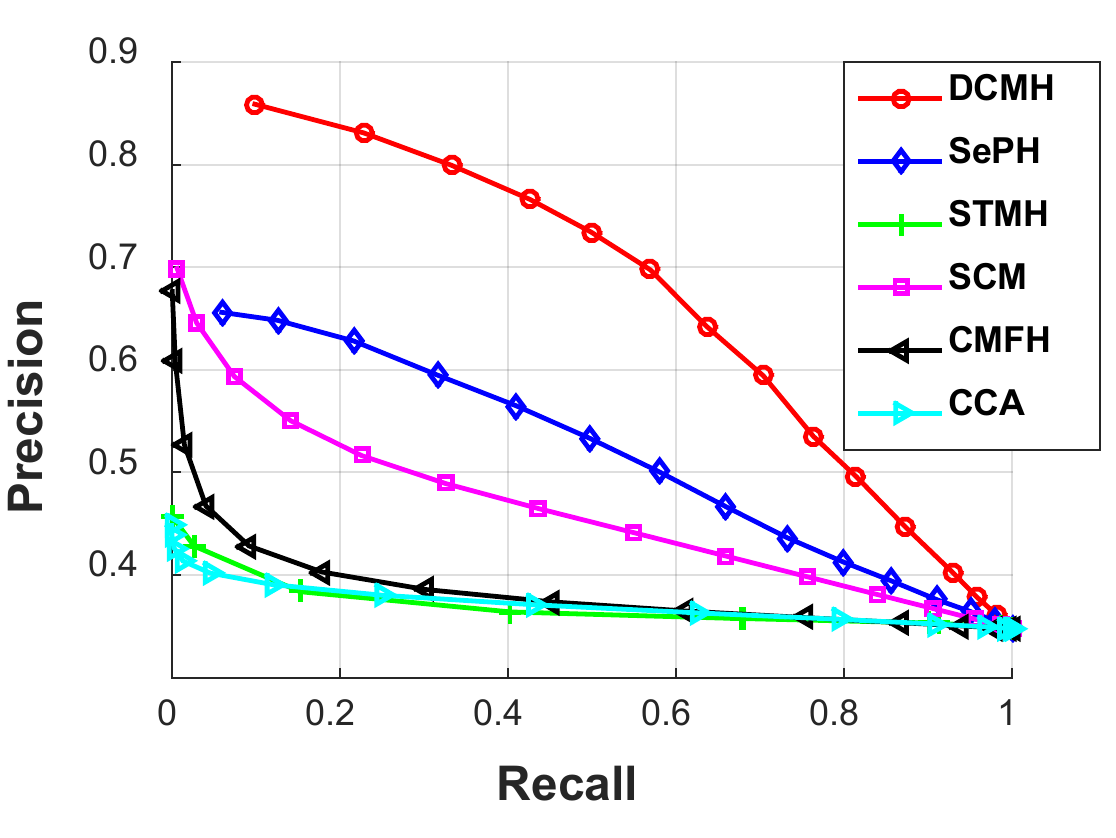}\\
    (d) Text $\to$ Image @NUS-WIDE
\end{minipage}\vspace*{-0pt}
\end{tabular}
\vspace{-0.2cm}
\caption{Precision-recall curves. The baselines are based on hand-crafted features. The code length is 16.}
\label{fig:mir_nus_pr_ori}
\end{figure*}

\begin{figure*}[t]
\centering
\small
\begin{tabular}{c@{ }@{ }c@{ }@{ }c@{ }@{ }c}
\begin{minipage}{0.23\linewidth}\centering
    \includegraphics[width=1\textwidth]{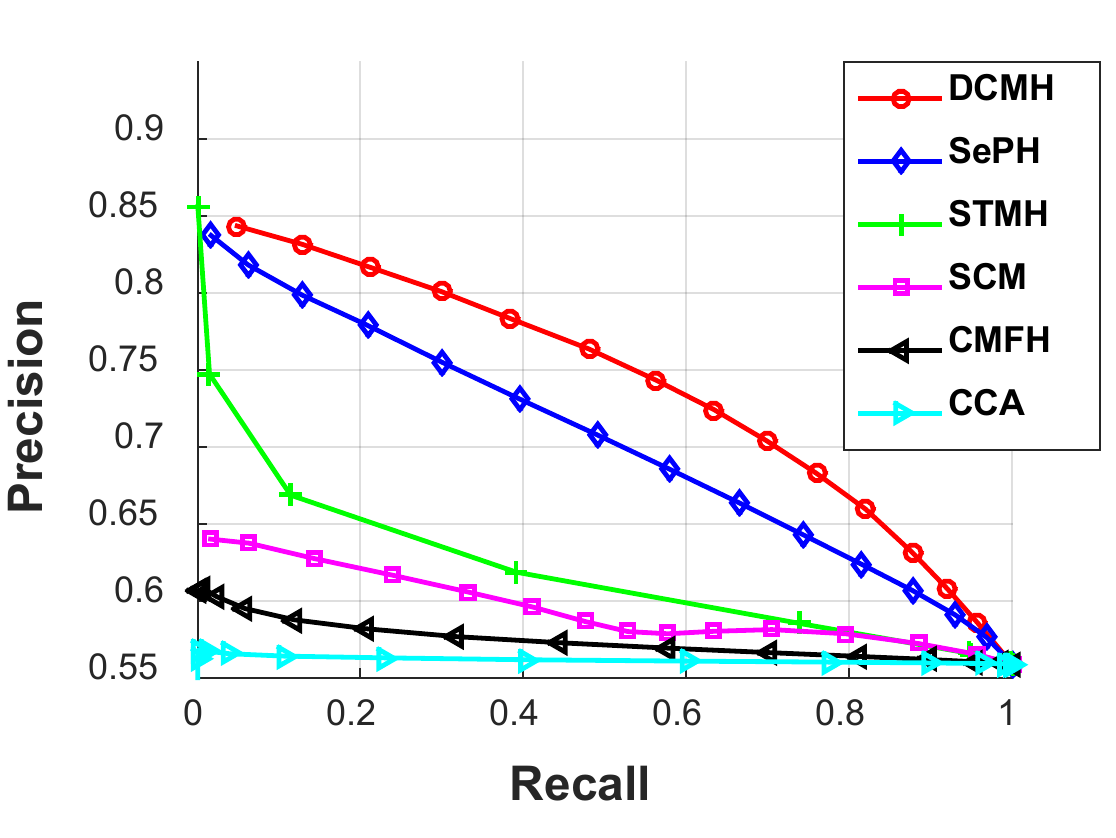}\\
    (a) Image $\to$ Text @MIRFLICKR-25K
\end{minipage} &
\begin{minipage}{0.23\linewidth}\centering
    \includegraphics[width=1\textwidth]{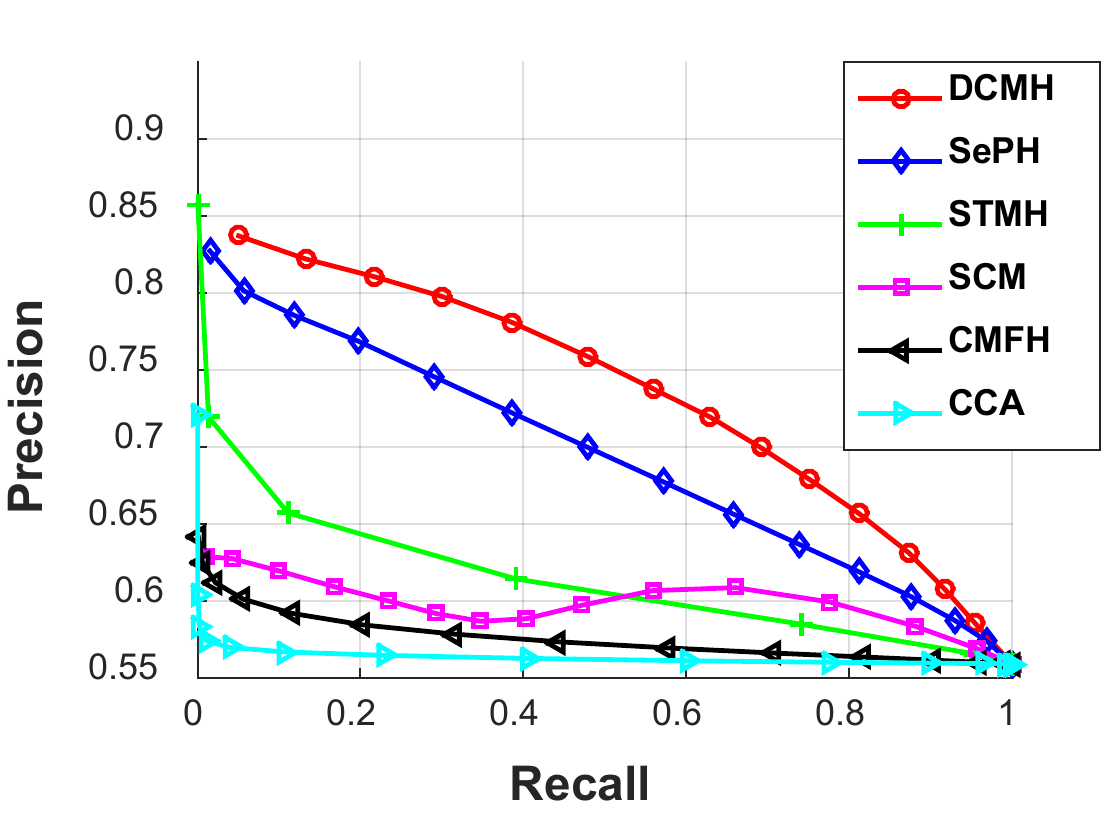}\\
    (b) Text $\to$ Image @MIRFLICKR-25K
\end{minipage} &
\begin{minipage}{0.23\linewidth}\centering
    \includegraphics[width=1\textwidth]{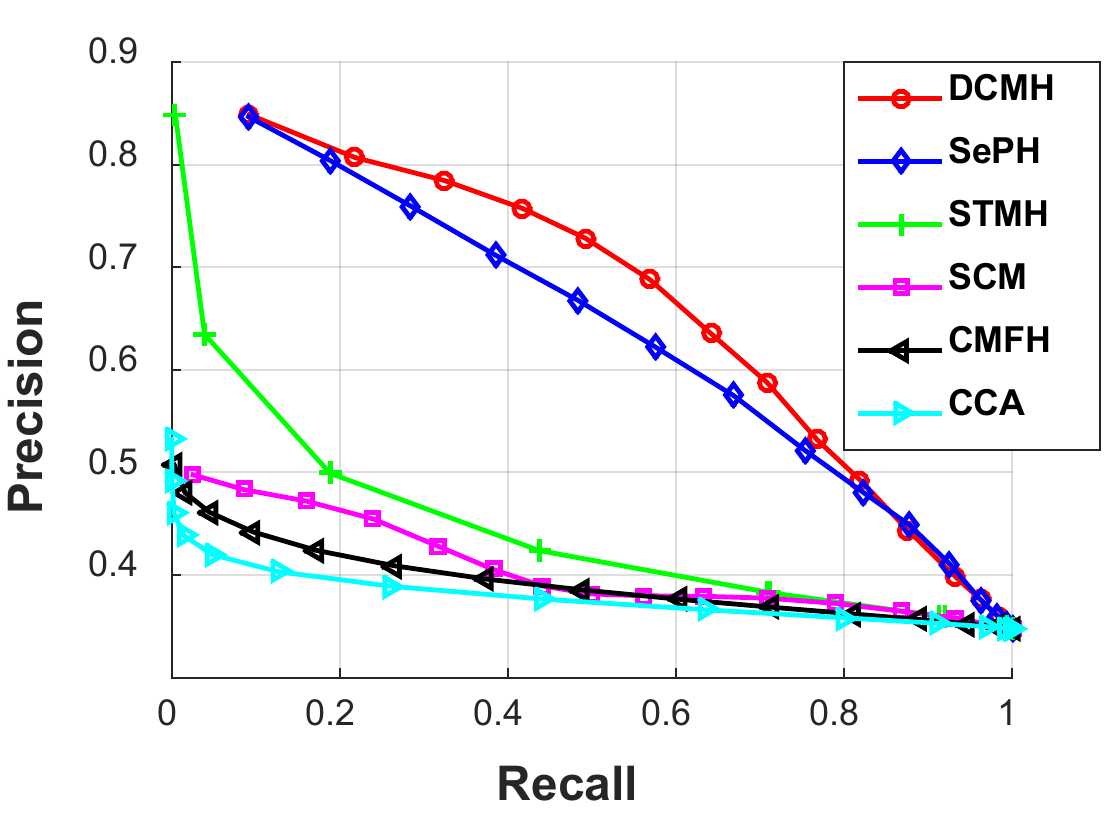}\\
    (c) Image $\to$ Text @NUS-WIDE
\end{minipage} &
\begin{minipage}{0.23\linewidth}\centering
    \includegraphics[width=1\textwidth]{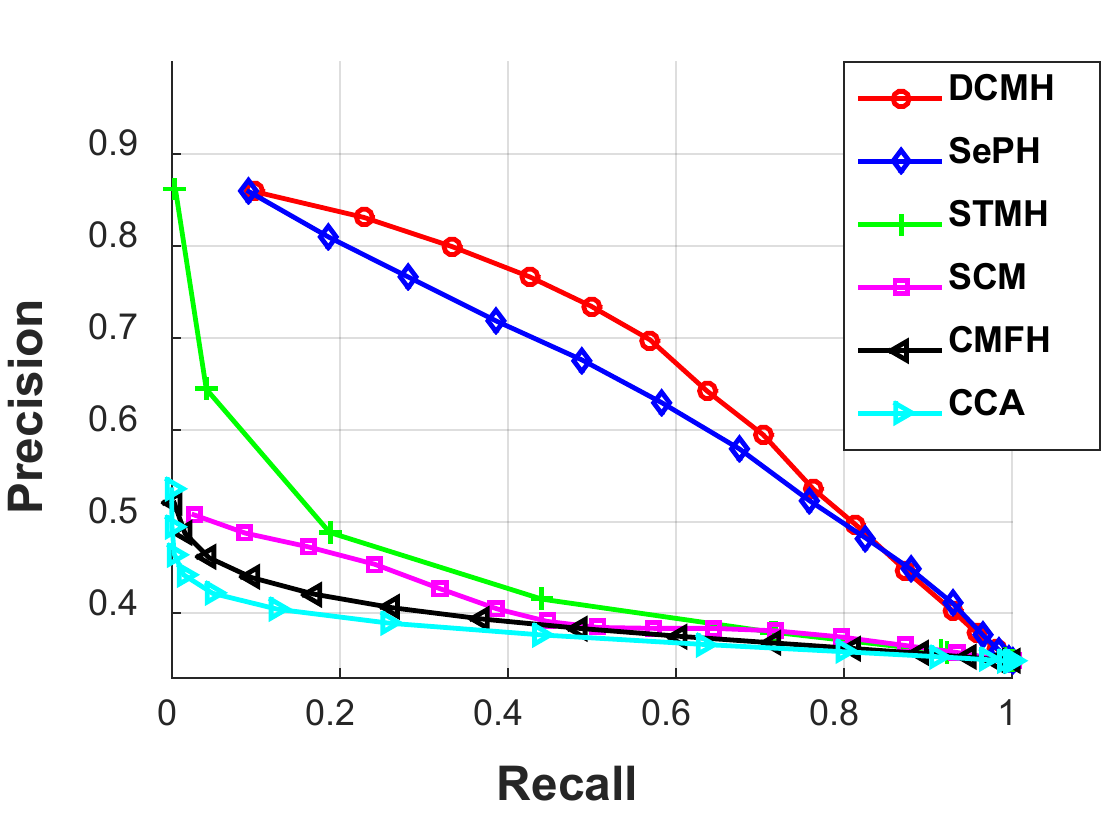}\\
    (d) Text $\to$ Image @NUS-WIDE
\end{minipage}\vspace*{-0pt}
\end{tabular}
\vspace{-0.2cm}
\caption{Precision-recall curves. The baselines are based on CNN-F features.  The code length is 16.}
\label{fig:mir_nus_pr_vgg}
\end{figure*}

\subsubsection{Hash Lookup}

In the hash lookup procedure, we can compute the precision, recall and F-measure for the returned points given any Hamming radius. The Hamming radius can take the values in $\{0,1,\dots,c\}$.  By varying the Hamming radius from 0 to $c$ with a stepsize 1, we can get the precision-recall curve.

Figure~\ref{fig:mir_nus_pr_ori} shows the precision-recall curve with code length 16 on two datasets, where the baselines use hand-drafted features. Here, ``Image $\to$ Text" denotes the case where the query is image and the database is text, and similar notations are used for other cases. We can find that DCMH can dramatically outperform the baselines.

Figure~\ref{fig:mir_nus_pr_vgg} shows the precision-recall curve with code length 16 on two datasets, where the baselines use CNN-F features. We can also find that DCMH can outperform all the other baselines with CNN-F features.

We select the best three methods and report their precision, recall and F-measure with Hamming radius $r=0,1,2$ in Table~\ref{table:mir_pr_vgg} on MIRFLICKR-25K when the code length is 16, where ``I" denotes image and ``T" denotes text. We can find that in all cases our DCMH can achieve the best recall and F-measure within Hamming radius $r=0,1,2$. For precision, DCMH outperforms SePH in all cases, but is outperformed by STMH in most cases. However, this does not mean that STMH is better than DCMH, because the recall of STMH is very poor. For example, assume there are 10,000 ground-truth similar points for a query on MIRFLICKR-25K. If we use an image query to retrieve text database with a Hamming radius 0, STMH only returns 3 points. However, our DCMH method can return nearly 580 points and 487 of them are ground-truth similar points. Hence, DCMH is more practical than STMH in real applications. From this perspective, F-measure is a more meaningful metric than precision and recall in the hash lookup procedure, and our DCMH achieves the best F-measure on all cases.

Please note that we only report the results when the code length is 16 due to space limitation. Our DCMH can also achieve the best performance on other cases with different number of code length. Furthermore, our DCMH is not sensitive to hyper-parameters $\gamma$ and $\eta$ when they are from the range $[0.5,2]$. All these experiments can be found in the supplementary materials.

\begin{table}[t]
\centering
\small
\caption{Precision, recall and F-measure on {{\emph{MIRFLICKR-25K}}} with CNN-F features. The best F-measure is shown in boldface.}
\label{table:mir_pr_vgg}
\begin{tabular}{|c|c|c|c|c|c|}
\hline
\multirow{2}{*}{Task} & \multirow{2}{*}{Method}&  \multirow{2}{*}{Metric}& \multicolumn{3}{|c|}{Hamming Radius} \\
\cline{4-6}
& &&$r=0$ &$r=1$ & $r=2$\\
\hline
\hline
\multirow{9}{*}{{I $\to$ T}}
& \multirow{3}{*}{DCMH} & Precision & 0.8434 & 0.8316 & 0.8166 \\
&                       & Recall    & 0.0487 & 0.1281 & 0.2121 \\
&                       & F-measure & \textbf{0.0920} & \textbf{0.2220} & \textbf{0.3367} \\
\cline{2-6}
& \multirow{3}{*}{SePH} & Precision & 0.8373 & 0.8182 & 0.7985 \\
&                       & Recall    & 0.0166 & 0.0620 & 0.1286 \\
&                       & F-measure & 0.0325 & {0.1153} & {0.2215} \\
\cline{2-6}
& \multirow{3}{*}{STMH} & Precision & 0.8560 & 0.8560 & 0.7473 \\
&                       & Recall    & 0.0003 & 0.0003 & 0.0147 \\
&                       & F-measure & 0.0007 & 0.0007 & 0.0287 \\
\hline
\multirow{9}{*}{{T $\to$ I}}
& \multirow{3}{*}{DCMH} & Precision & 0.8370 & 0.8220 & 0.8106 \\
&                       & Recall    & 0.0504 & 0.1329 & 0.2164 \\
&                       & F-measure & \textbf{0.0951} & \textbf{0.2288} & \textbf{0.3416} \\
\cline{2-6}
& \multirow{3}{*}{SePH} & Precision & 0.8273 & 0.8014 & 0.7853 \\
&                       & Recall    & 0.0151 & 0.0566 & 0.1193 \\
&                       & F-measure & 0.0297 & 0.1058 & 0.2072 \\
\cline{2-6}
& \multirow{3}{*}{STMH} & Precision & 0.8578 & 0.8578 & 0.7194 \\
&                       & Recall    & 0.0003 & 0.0003 & 0.0134 \\
&                       & F-measure & 0.0006 & 0.0006 & 0.0264 \\
\hline
\end{tabular}
\vspace{-0.5cm}
\end{table}

%Table~\ref{table:mir_pr_ori} and
%Table~\ref{table:nus_pr_ori} and Table~\ref{table:nus_pr_vgg} report the precision-recall on NUS-WIDE dataset.

%\subsection{Sensitivity to Hyper-Parameters}

\section{Conclusion}
\label{sec:conclusion}
In this paper, we have proposed a novel hashing method, called DCMH, for cross-modal retrieval applications.
\mbox{DCMH} is an end-to-end learning framework which can perform feature learning from scratch. To the best of our knowledge, DCMH is the first cross-modal hashing method which can perform simultaneous feature learning and hash-code learning in the same framework.
Experiments on two datasets show that DCMH can outperform other baselines to achieve the state-of-the-art performance in real applications.
\bibliography{reference}
\bibliographystyle{icml2016}

\appendix
%\section{Precision-Recall Curve}
\begin{figure*}[t]
\centering
\small
\begin{tabular}{c@{ }@{ }c@{ }@{ }c@{ }@{ }c}
\begin{minipage}{0.23\linewidth}\centering
    \includegraphics[width=1\textwidth]{MIR_i2t_ORI_16.pdf}\\
    (a) Image $\to$ Text @MIRFLICKR-25K
\end{minipage} &
\begin{minipage}{0.23\linewidth}\centering
    \includegraphics[width=1\textwidth]{MIR_t2i_ORI_16.pdf}\\
    (b) Text $\to$ Image @MIRFLICKR-25K
\end{minipage} &
\begin{minipage}{0.23\linewidth}\centering
    \includegraphics[width=1\textwidth]{NUS_i2t_ORI_16.pdf}\\
    (c) Image $\to$ Text @NUS-WIDE
\end{minipage} &
\begin{minipage}{0.23\linewidth}\centering
    \includegraphics[width=1\textwidth]{NUS_t2i_ORI_16.pdf}\\
    (d) Text $\to$ Image @NUS-WIDE
\end{minipage}\vspace*{-0pt}
\end{tabular}
\begin{tabular}{c@{ }@{ }c@{ }@{ }c@{ }@{ }c}
\begin{minipage}{0.23\linewidth}\centering
    \includegraphics[width=1\textwidth]{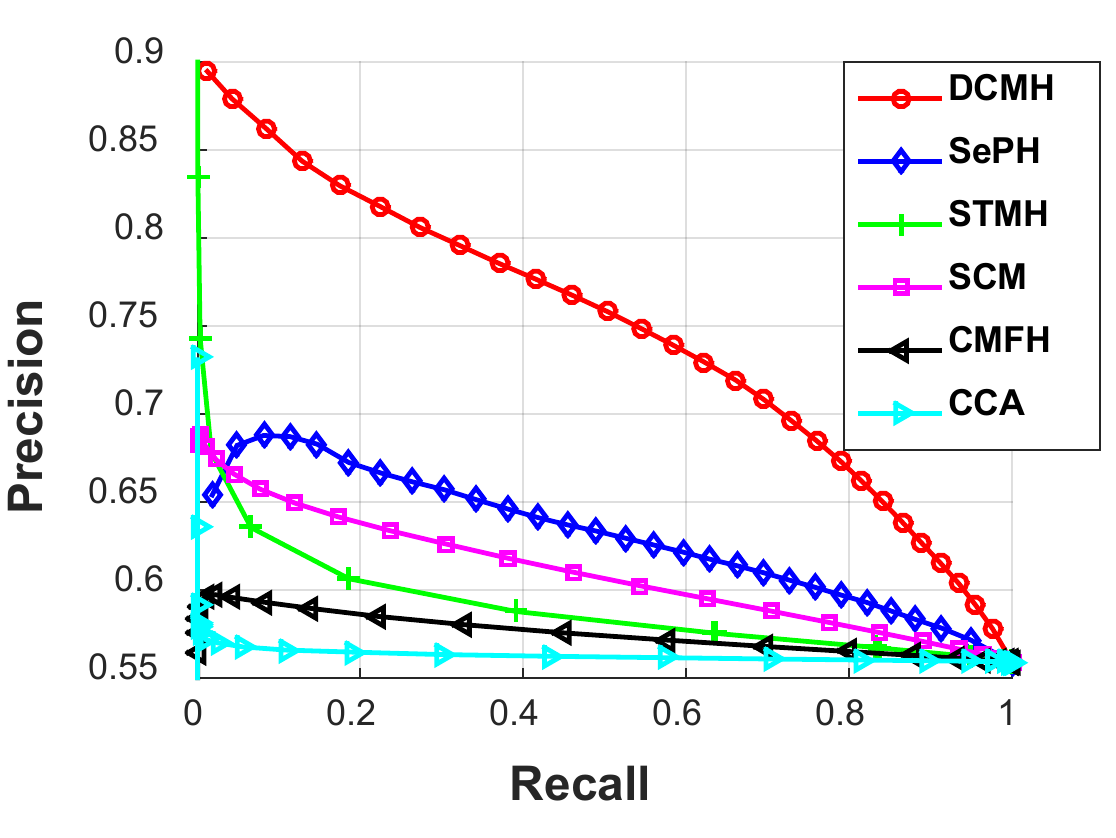}\\
    (a) Image $\to$ Text @MIRFLICKR-25K
\end{minipage} &
\begin{minipage}{0.23\linewidth}\centering
    \includegraphics[width=1\textwidth]{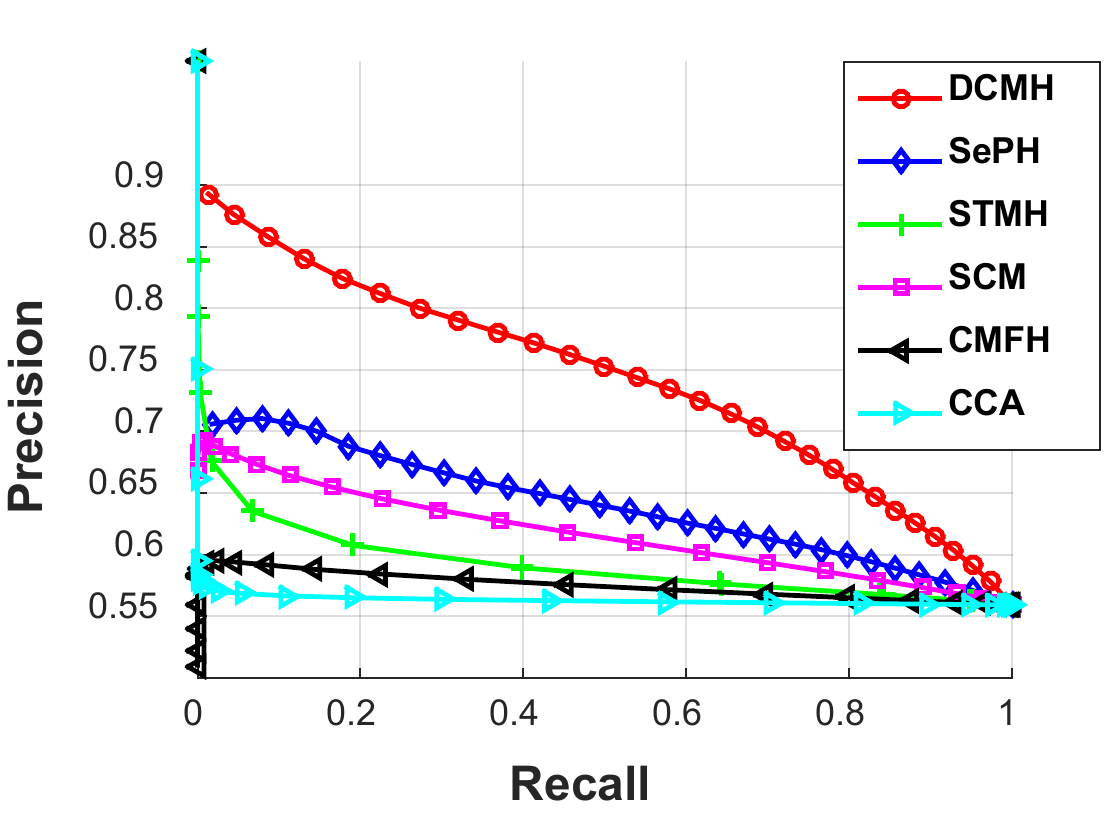}\\
    (b) Text $\to$ Image @MIRFLICKR-25K
\end{minipage} &
\begin{minipage}{0.23\linewidth}\centering
    \includegraphics[width=1\textwidth]{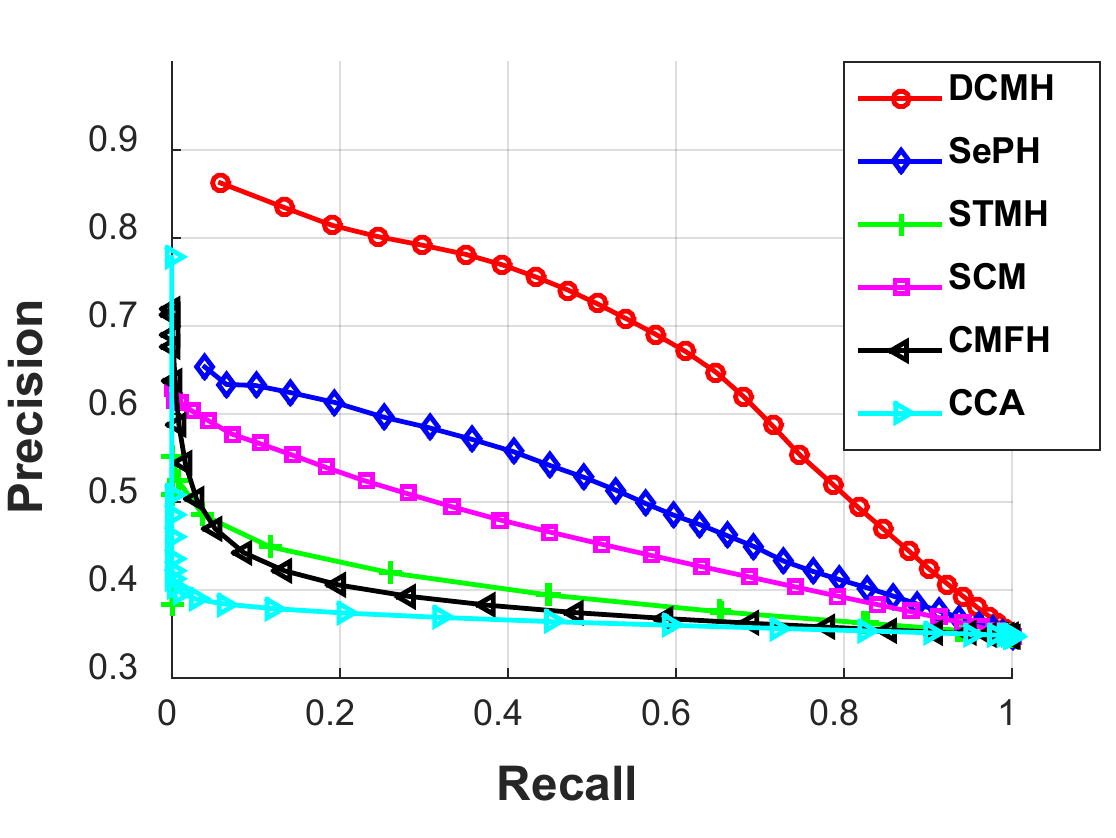}\\
    (c) Image $\to$ Text @NUS-WIDE
\end{minipage} &
\begin{minipage}{0.23\linewidth}\centering
    \includegraphics[width=1\textwidth]{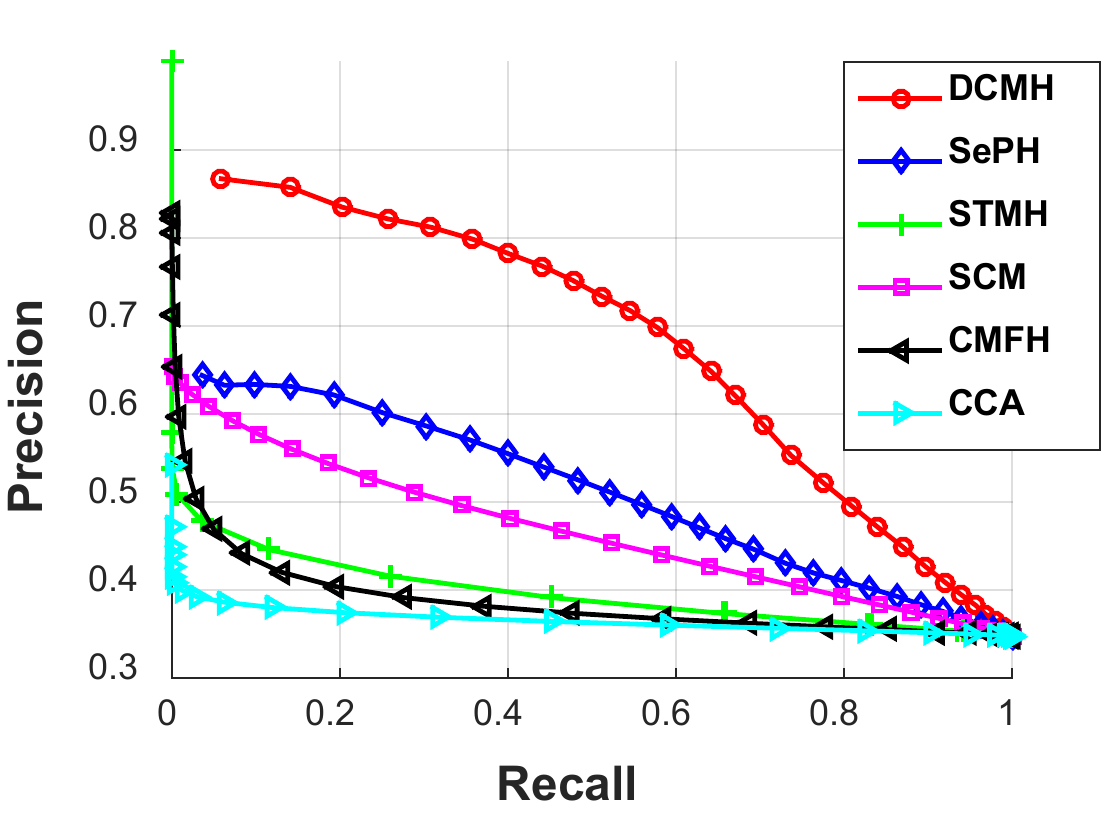}\\
    (d) Text $\to$ Image @NUS-WIDE
\end{minipage}\vspace*{-0pt}
\end{tabular}
\begin{tabular}{c@{ }@{ }c@{ }@{ }c@{ }@{ }c}
\begin{minipage}{0.23\linewidth}\centering
    \includegraphics[width=1\textwidth]{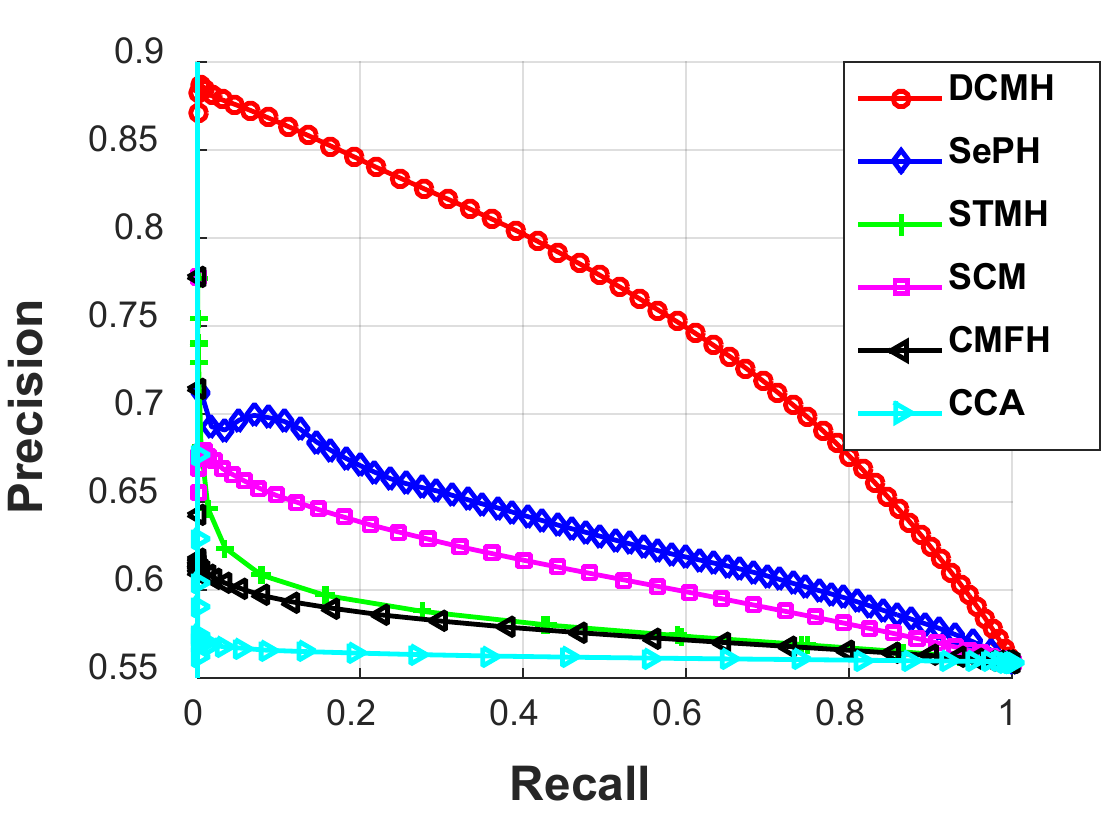}\\
    (a) Image $\to$ Text @MIRFLICKR-25K
\end{minipage} &
\begin{minipage}{0.23\linewidth}\centering
    \includegraphics[width=1\textwidth]{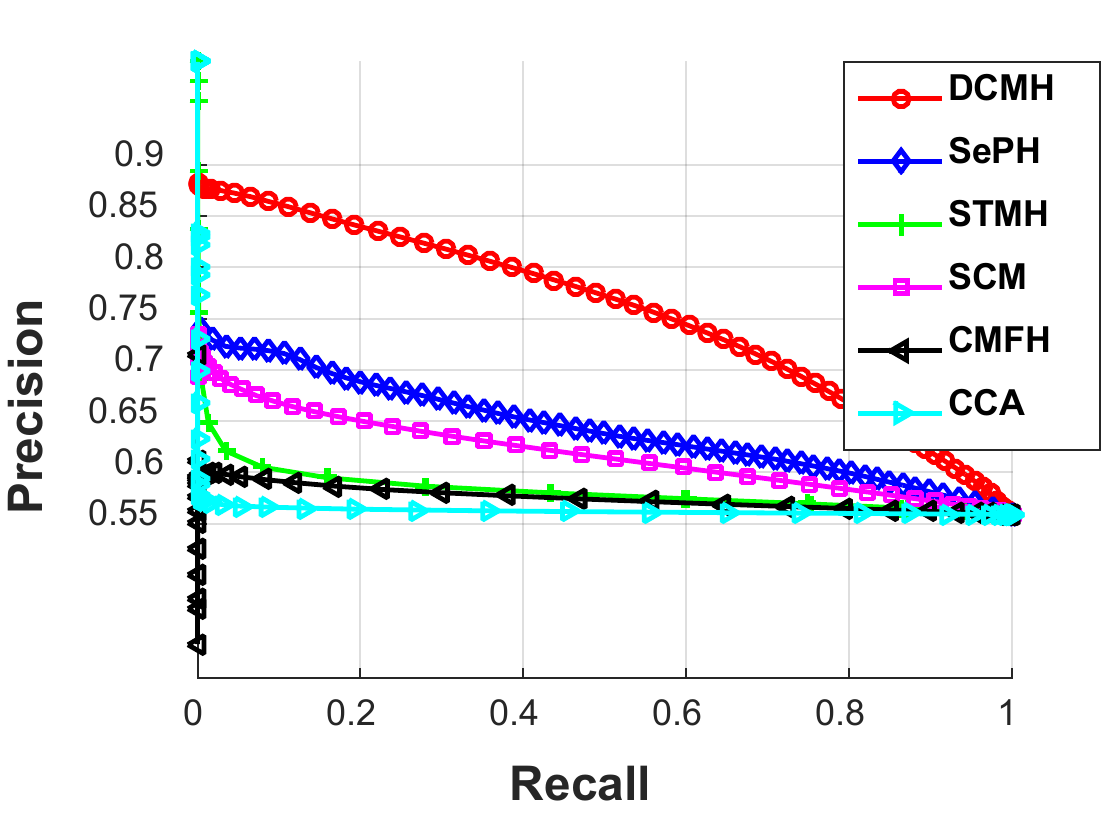}\\
    (b) Text $\to$ Image @MIRFLICKR-25K
\end{minipage} &
\begin{minipage}{0.23\linewidth}\centering
    \includegraphics[width=1\textwidth]{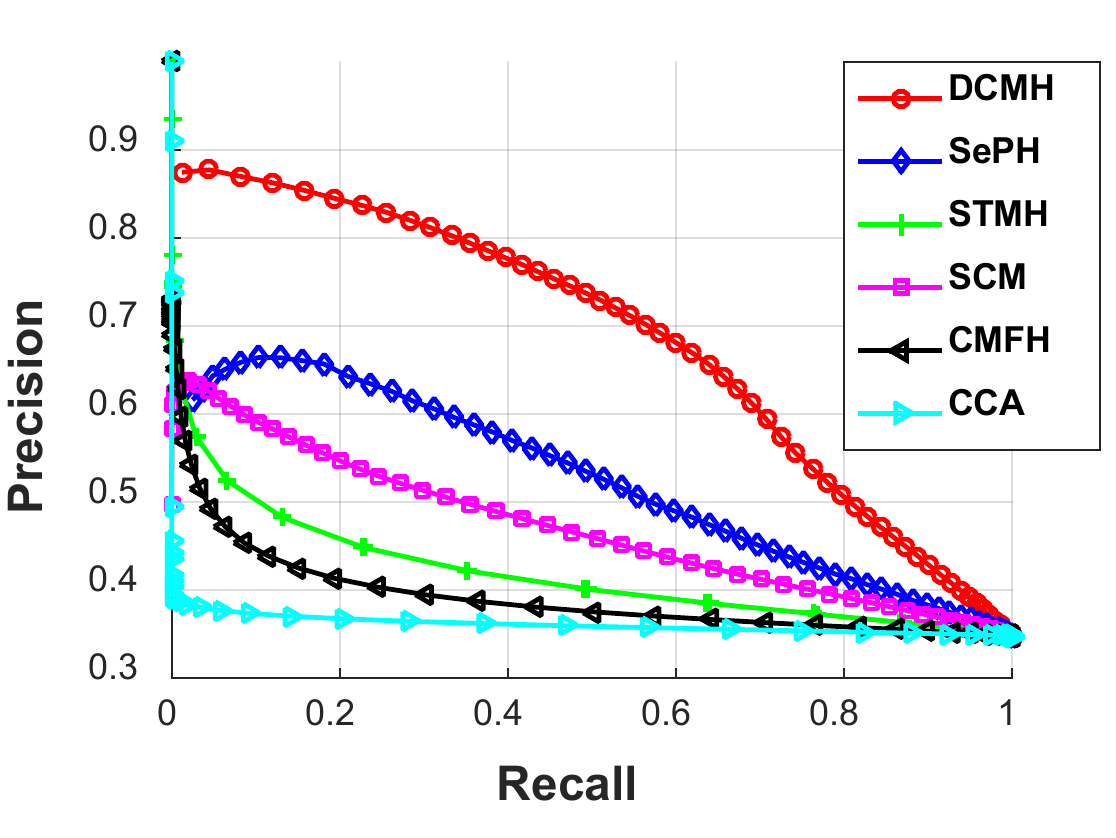}\\
    (c) Image $\to$ Text @NUS-WIDE
\end{minipage} &
\begin{minipage}{0.23\linewidth}\centering
    \includegraphics[width=1\textwidth]{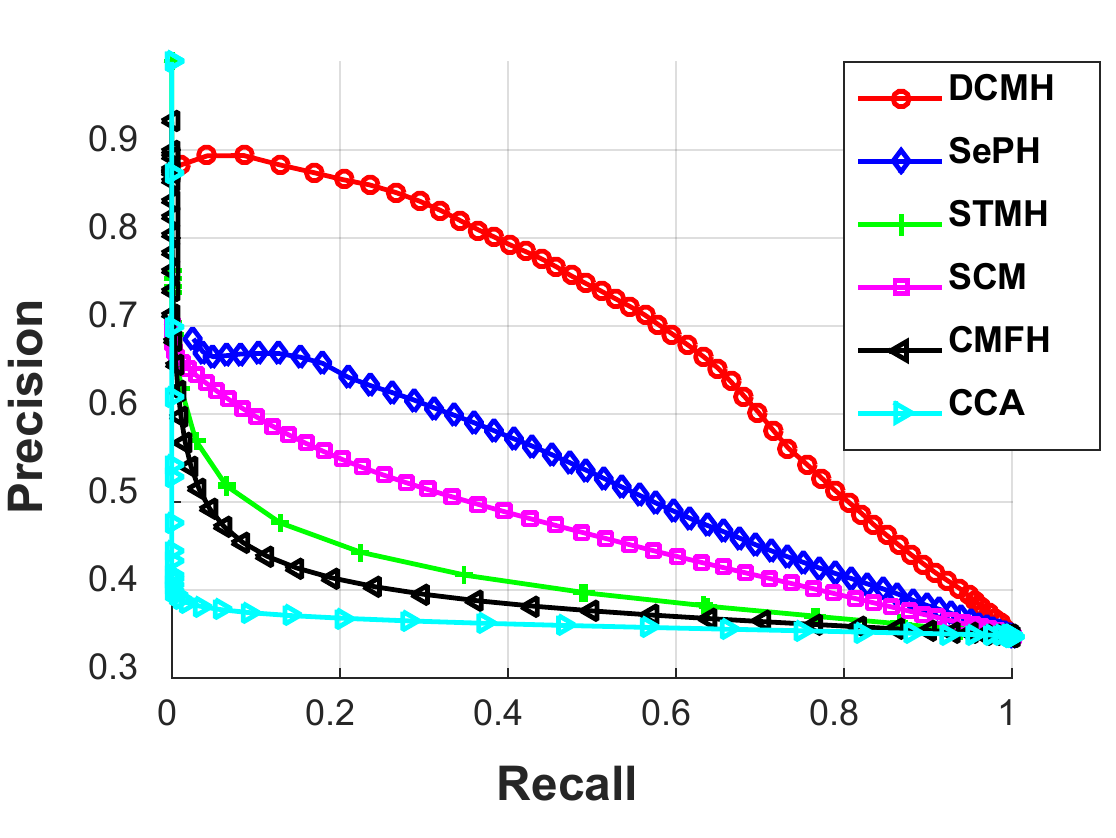}\\
    (d) Text $\to$ Image @NUS-WIDE
\end{minipage}\vspace*{-0pt}
\end{tabular}
\vspace{-0.2cm}
\caption{Precision-recall curves. The baselines are based on hand-crafted features. The first row is for 16 bits, the second row is for 32 bits, and the third row is for 64 bits.}
\label{fig:mir_two_pr_ori}
\end{figure*}

\begin{figure*}[t]
\centering
\small
\begin{tabular}{c@{ }@{ }c@{ }@{ }c@{ }@{ }c}
\begin{minipage}{0.23\linewidth}\centering
    \includegraphics[width=1\textwidth]{MIR_i2t_VGG_16.pdf}\\
    (a) Image $\to$ Text @MIRFLICKR-25K
\end{minipage} &
\begin{minipage}{0.23\linewidth}\centering
    \includegraphics[width=1\textwidth]{MIR_t2i_VGG_16.pdf}\\
    (b) Text $\to$ Image @MIRFLICKR-25K
\end{minipage} &
\begin{minipage}{0.23\linewidth}\centering
    \includegraphics[width=1\textwidth]{NUS_i2t_VGG_16.pdf}\\
    (c) Image $\to$ Text @NUS-WIDE
\end{minipage} &
\begin{minipage}{0.23\linewidth}\centering
    \includegraphics[width=1\textwidth]{NUS_t2i_VGG_16.pdf}\\
    (d) Text $\to$ Image @NUS-WIDE
\end{minipage}\vspace*{-0pt}
\end{tabular}
\begin{tabular}{c@{ }@{ }c@{ }@{ }c@{ }@{ }c}
\begin{minipage}{0.23\linewidth}\centering
    \includegraphics[width=1\textwidth]{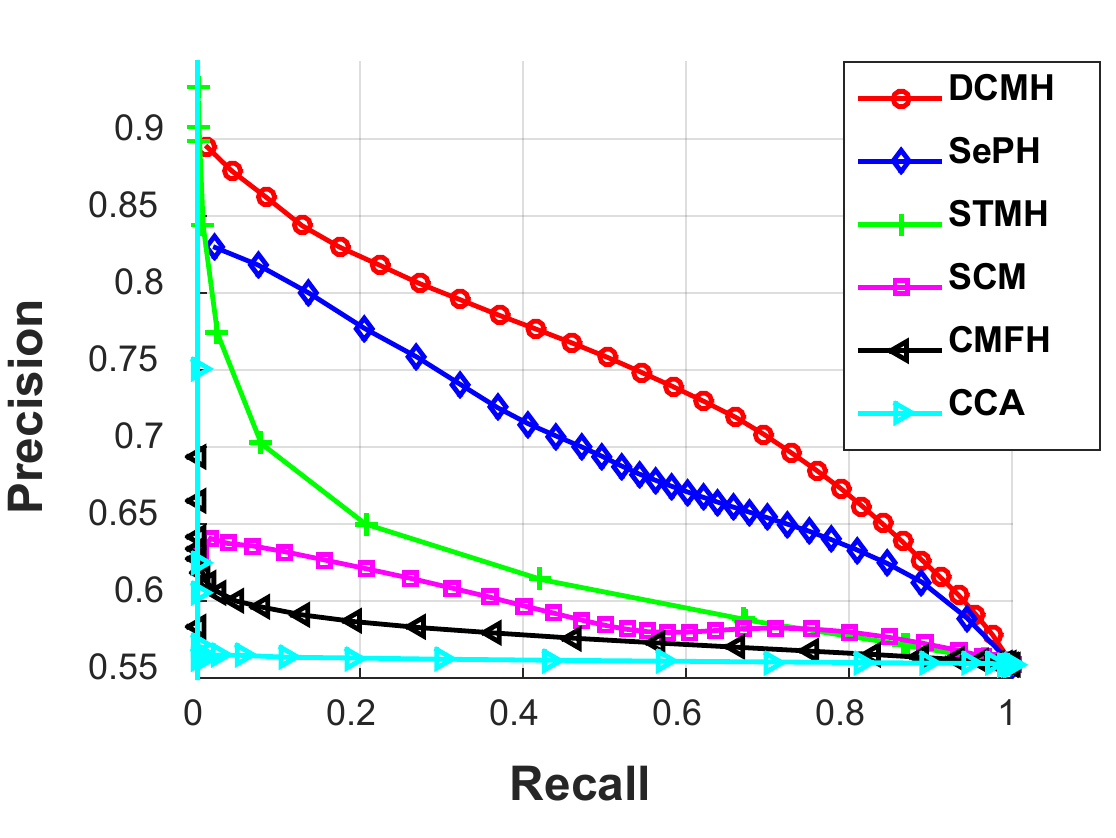}\\
    (a) Image $\to$ Text @MIRFLICKR-25K
\end{minipage} &
\begin{minipage}{0.23\linewidth}\centering
    \includegraphics[width=1\textwidth]{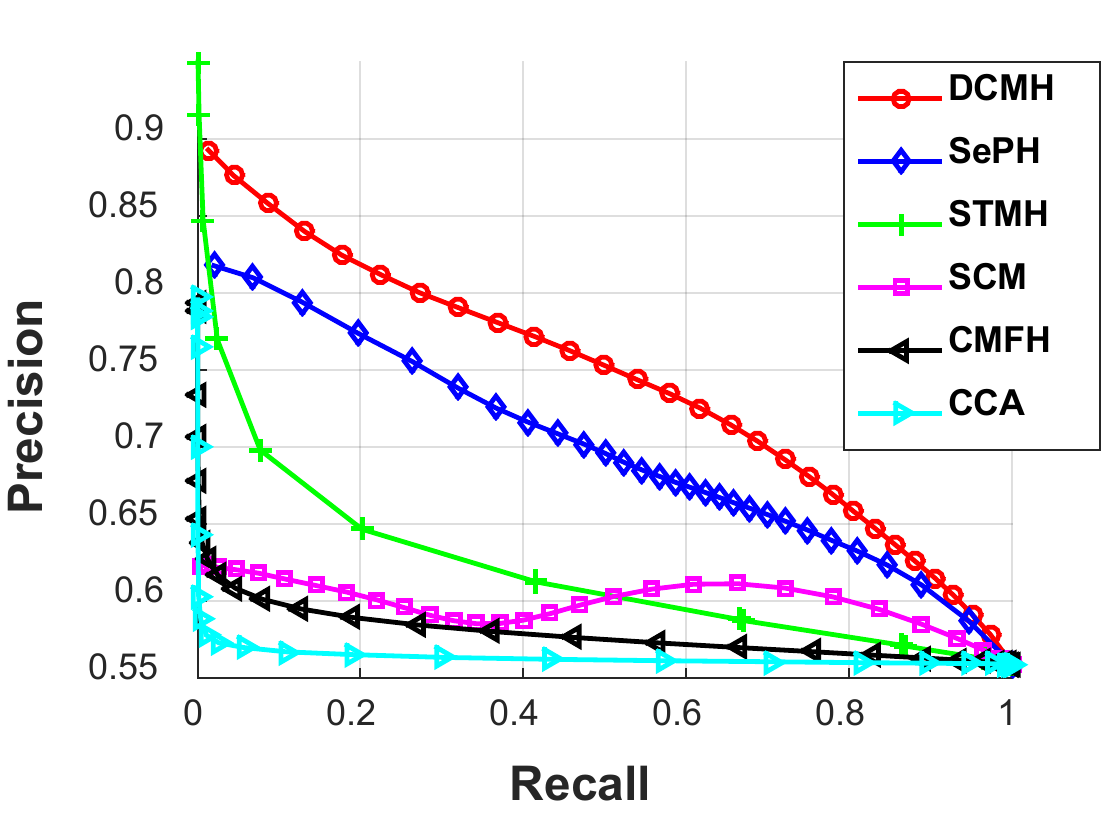}\\
    (b) Text $\to$ Image @MIRFLICKR-25K
\end{minipage} &
\begin{minipage}{0.23\linewidth}\centering
    \includegraphics[width=1\textwidth]{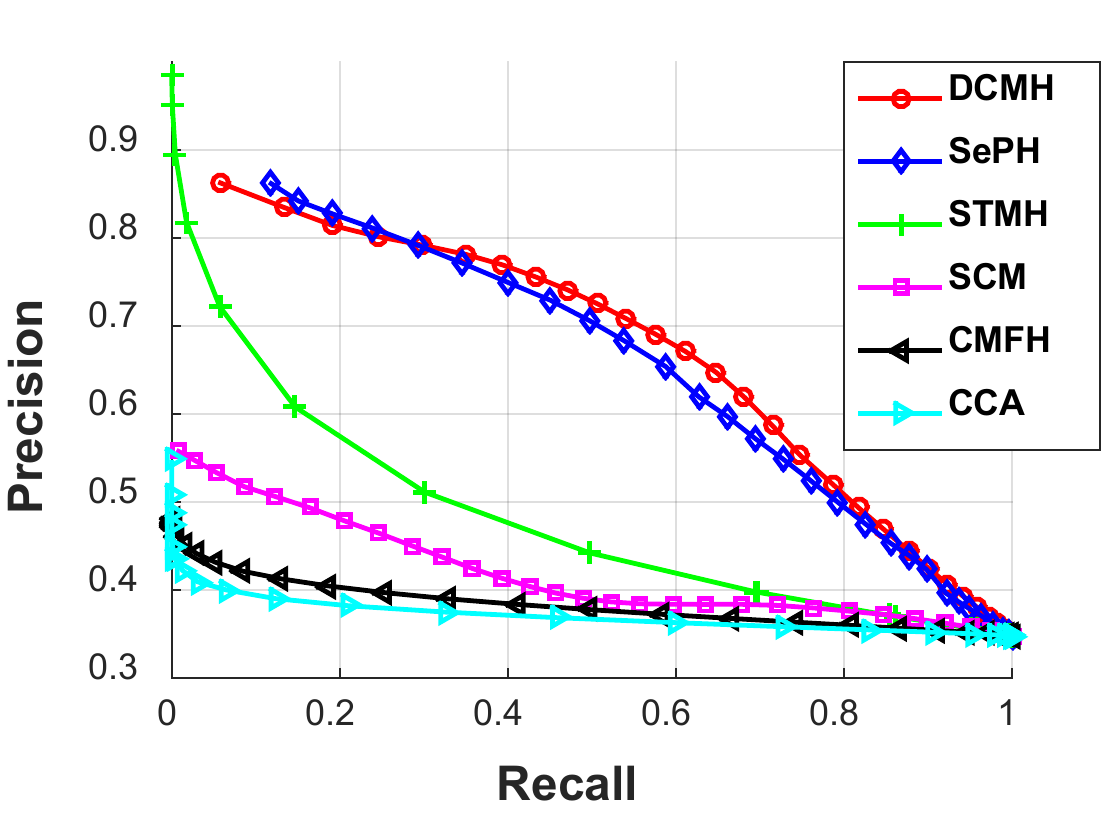}\\
    (c) Image $\to$ Text @NUS-WIDE
\end{minipage} &
\begin{minipage}{0.23\linewidth}\centering
    \includegraphics[width=1\textwidth]{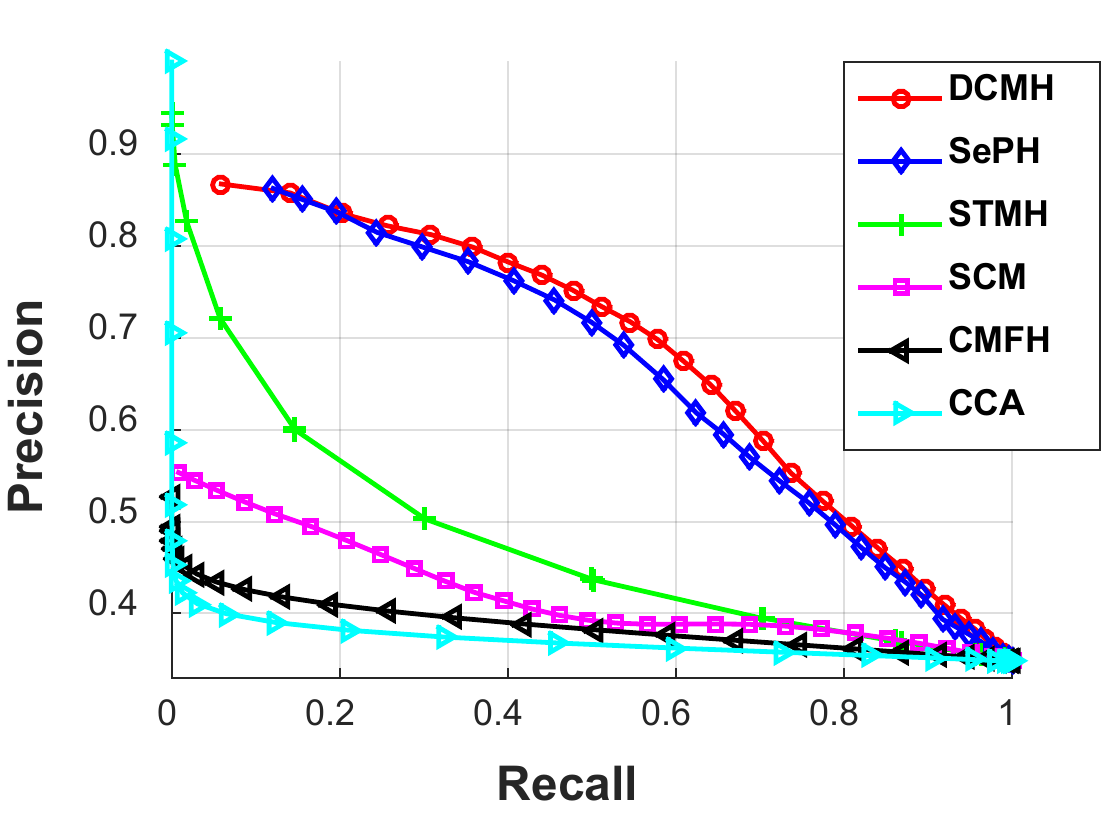}\\
    (d) Text $\to$ Image @NUS-WIDE
\end{minipage}\vspace*{-0pt}
\end{tabular}
\begin{tabular}{c@{ }@{ }c@{ }@{ }c@{ }@{ }c}
\begin{minipage}{0.23\linewidth}\centering
    \includegraphics[width=1\textwidth]{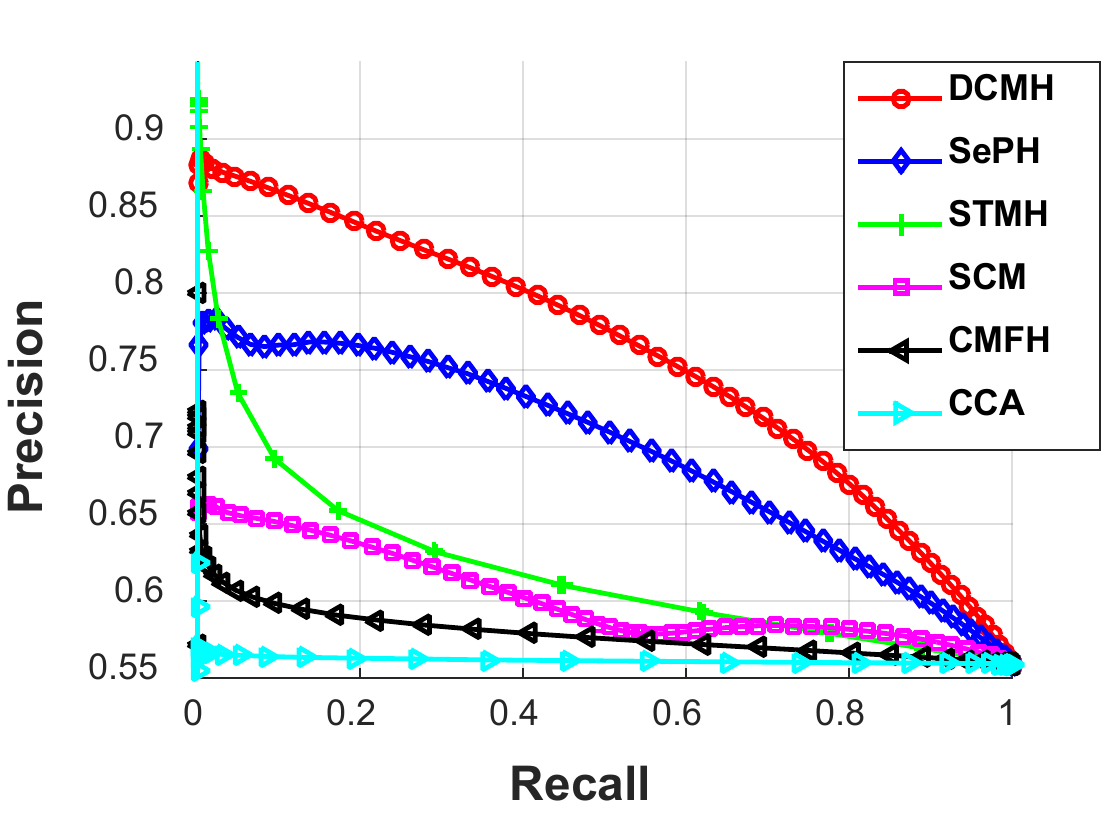}\\
    (a) Image $\to$ Text @MIRFLICKR-25K
\end{minipage} &
\begin{minipage}{0.23\linewidth}\centering
    \includegraphics[width=1\textwidth]{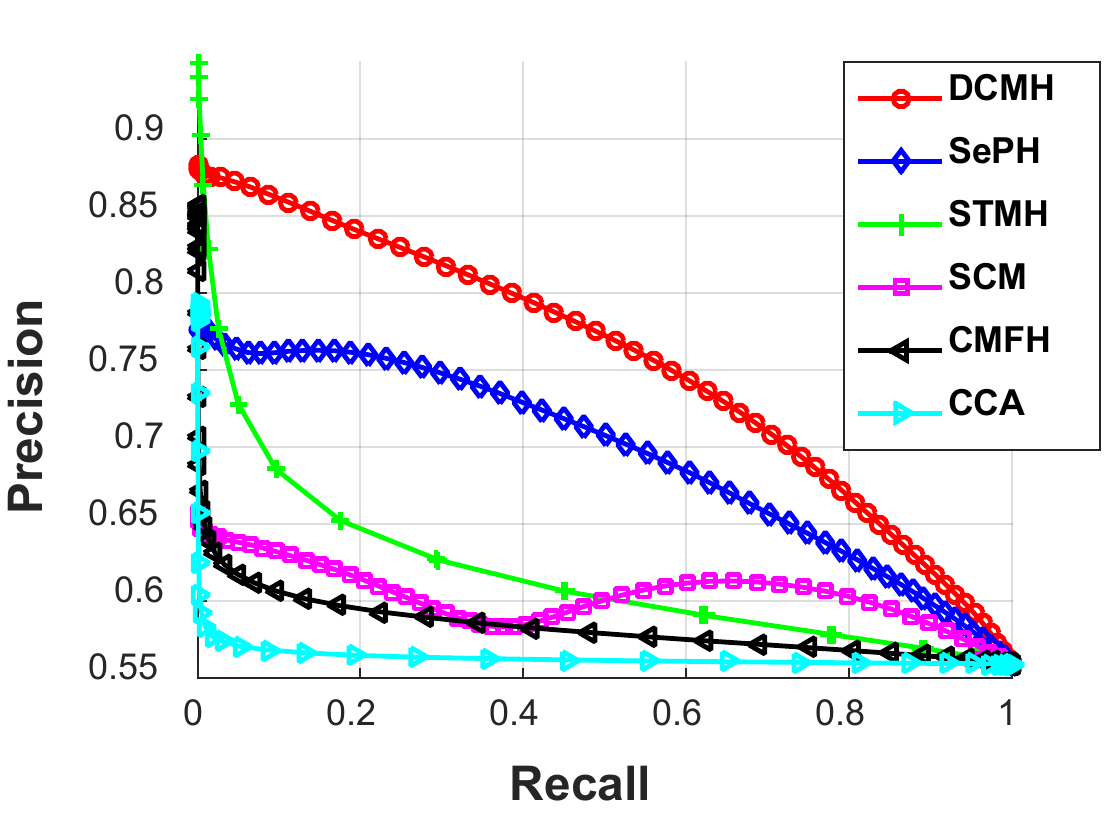}\\
    (b) Text $\to$ Image @MIRFLICKR-25K
\end{minipage} &
\begin{minipage}{0.23\linewidth}\centering
    \includegraphics[width=1\textwidth]{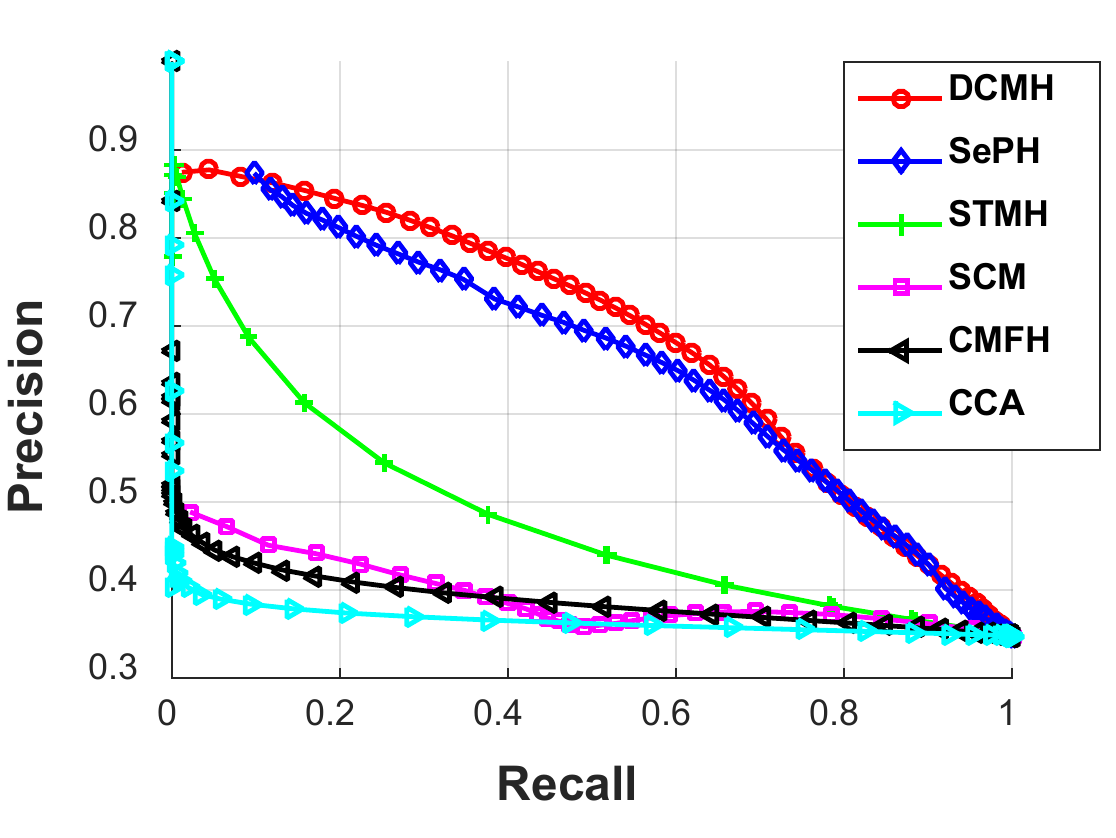}\\
    (c) Image $\to$ Text @NUS-WIDE
\end{minipage} &
\begin{minipage}{0.23\linewidth}\centering
    \includegraphics[width=1\textwidth]{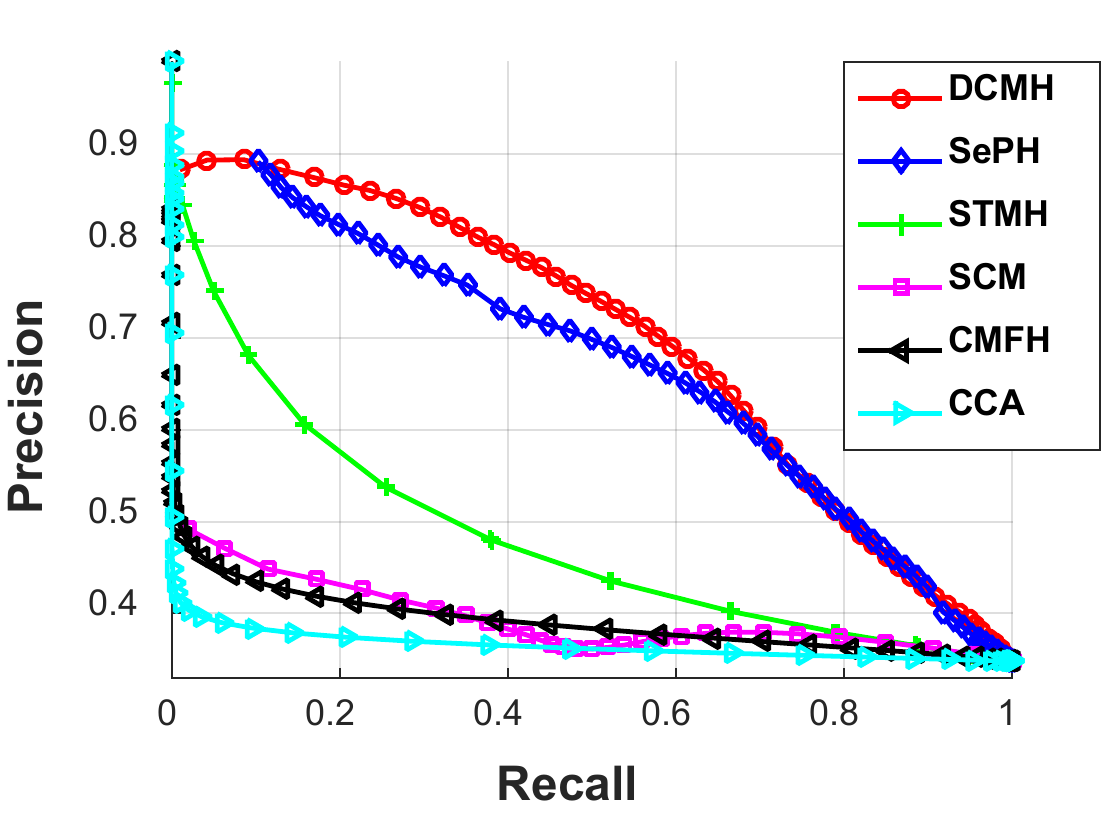}\\
    (d) Text $\to$ Image @NUS-WIDE
\end{minipage}\vspace*{-0pt}
\end{tabular}
\vspace{-0.2cm}
\caption{Precision-recall curves. The baselines are based on CNN-F features. The first row is for 16 bits, the second row is for 32 bits, and the third row is for 64 bits.}
\label{fig:mir_two_pr_vgg}
\end{figure*}

\begin{figure*}[t]
\centering
\small
\begin{tabular}{c@{ }@{ }c@{ }@{ }c@{ }@{ }c}
\begin{minipage}{0.23\linewidth}\centering
    \includegraphics[width=1\textwidth]{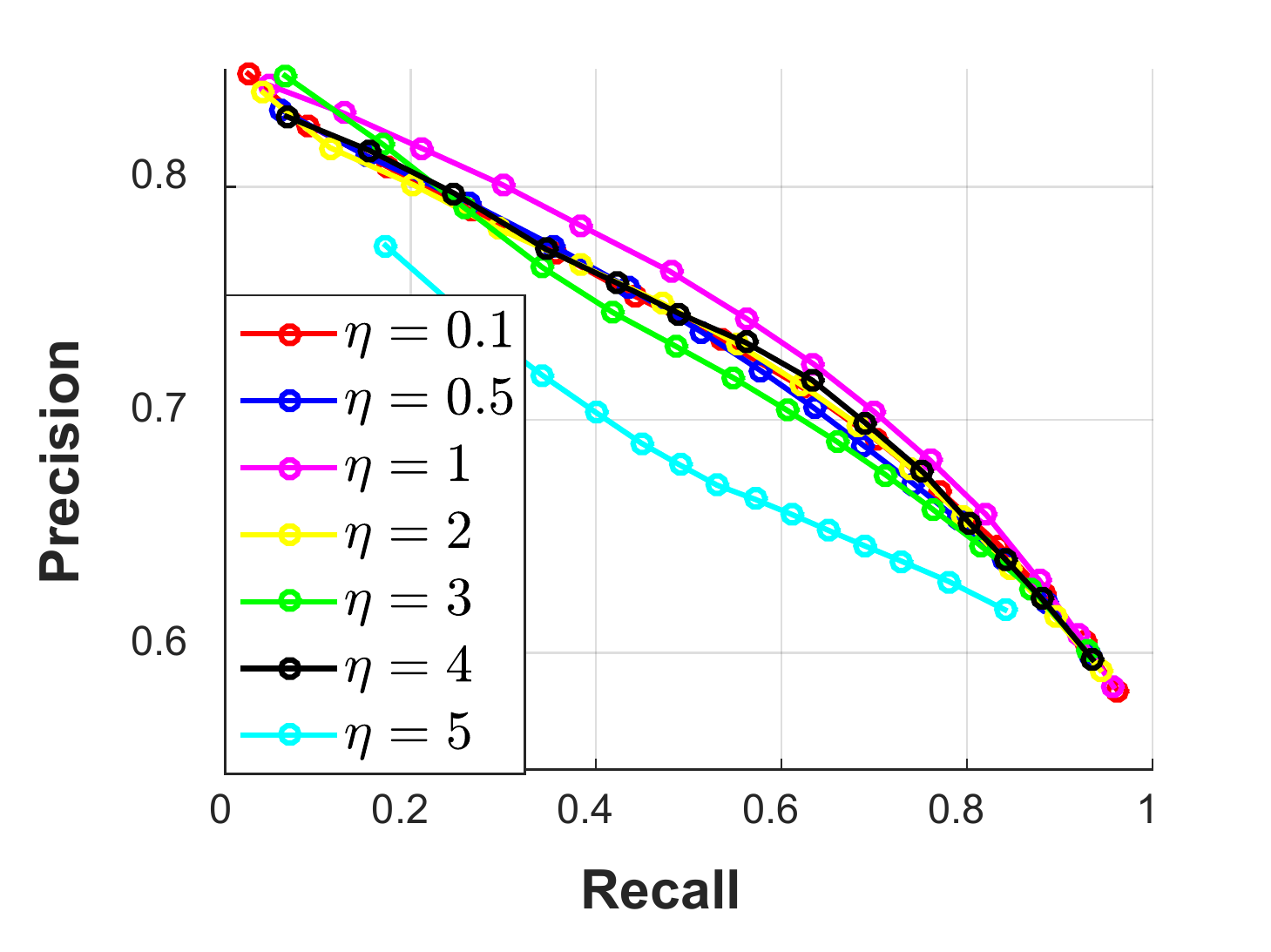}\\
    (a) Image $\to$ Text @MIRFLICKR-25K
\end{minipage} &
\begin{minipage}{0.23\linewidth}\centering
    \includegraphics[width=1\textwidth]{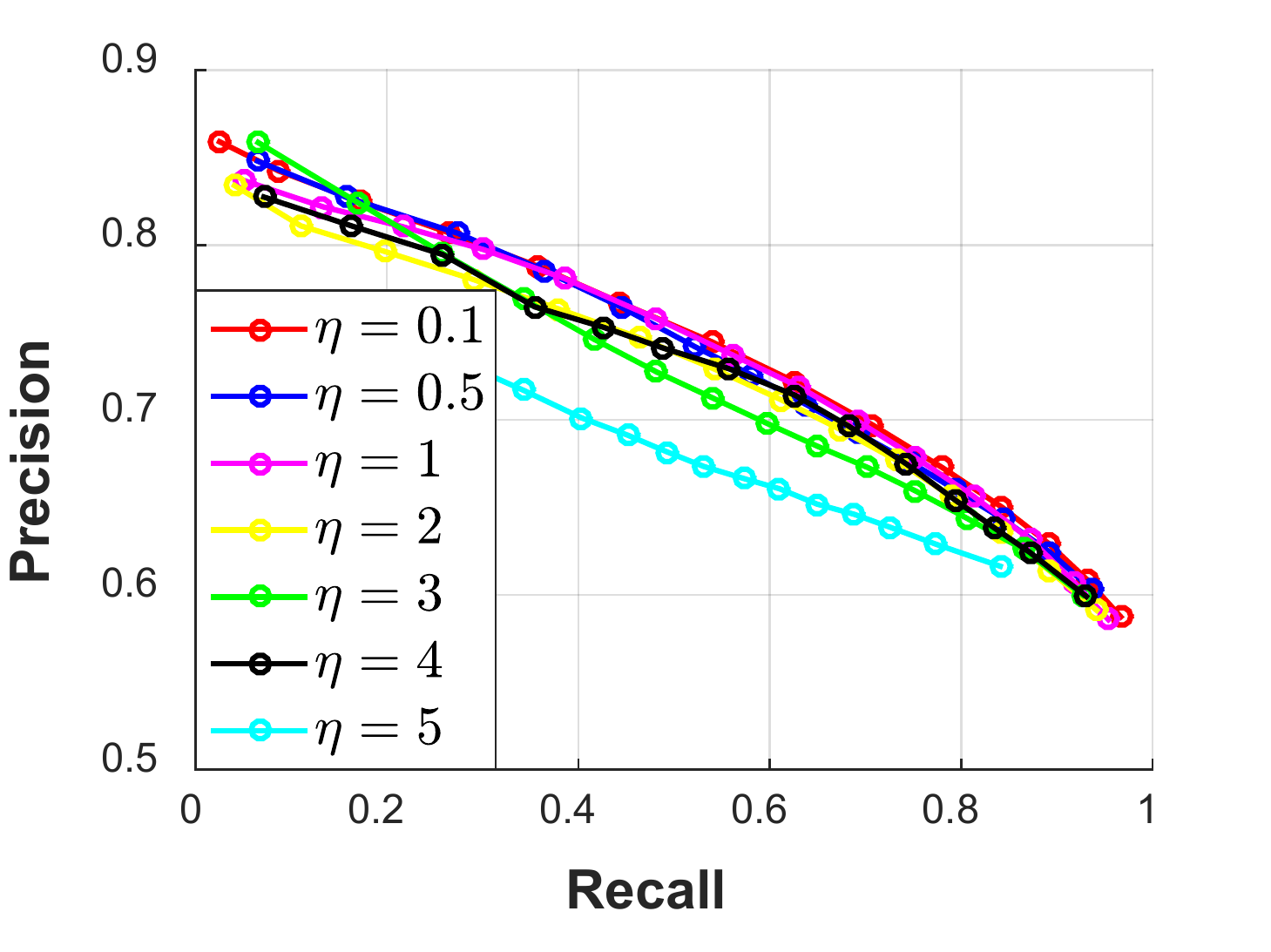}\\
    (b) Text $\to$ Image @MIRFLICKR-25K
\end{minipage} &
\begin{minipage}{0.23\linewidth}\centering
    \includegraphics[width=1\textwidth]{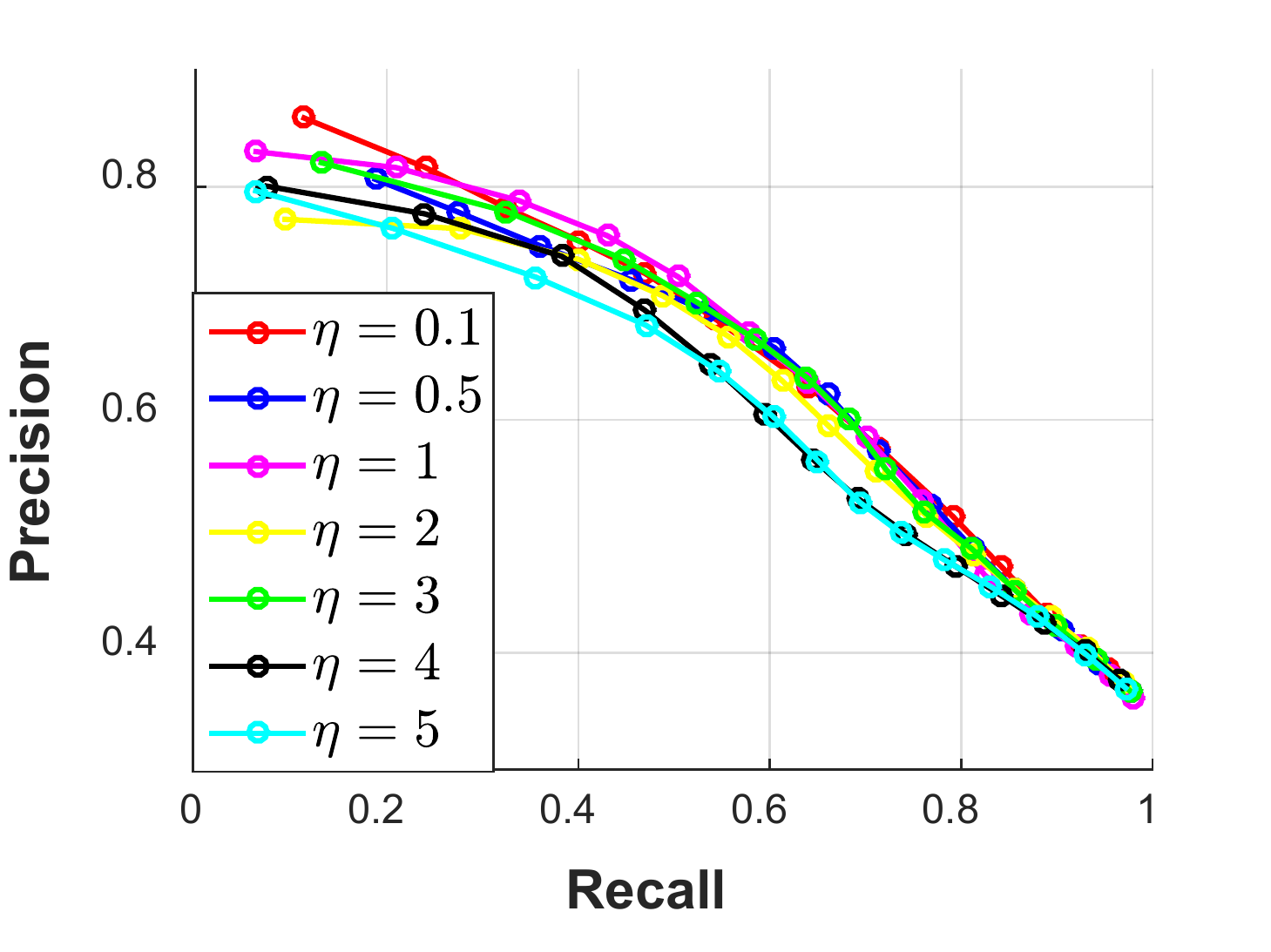}\\
    (c) Image $\to$ Text @NUS-WIDE
\end{minipage} &
\begin{minipage}{0.23\linewidth}\centering
    \includegraphics[width=1\textwidth]{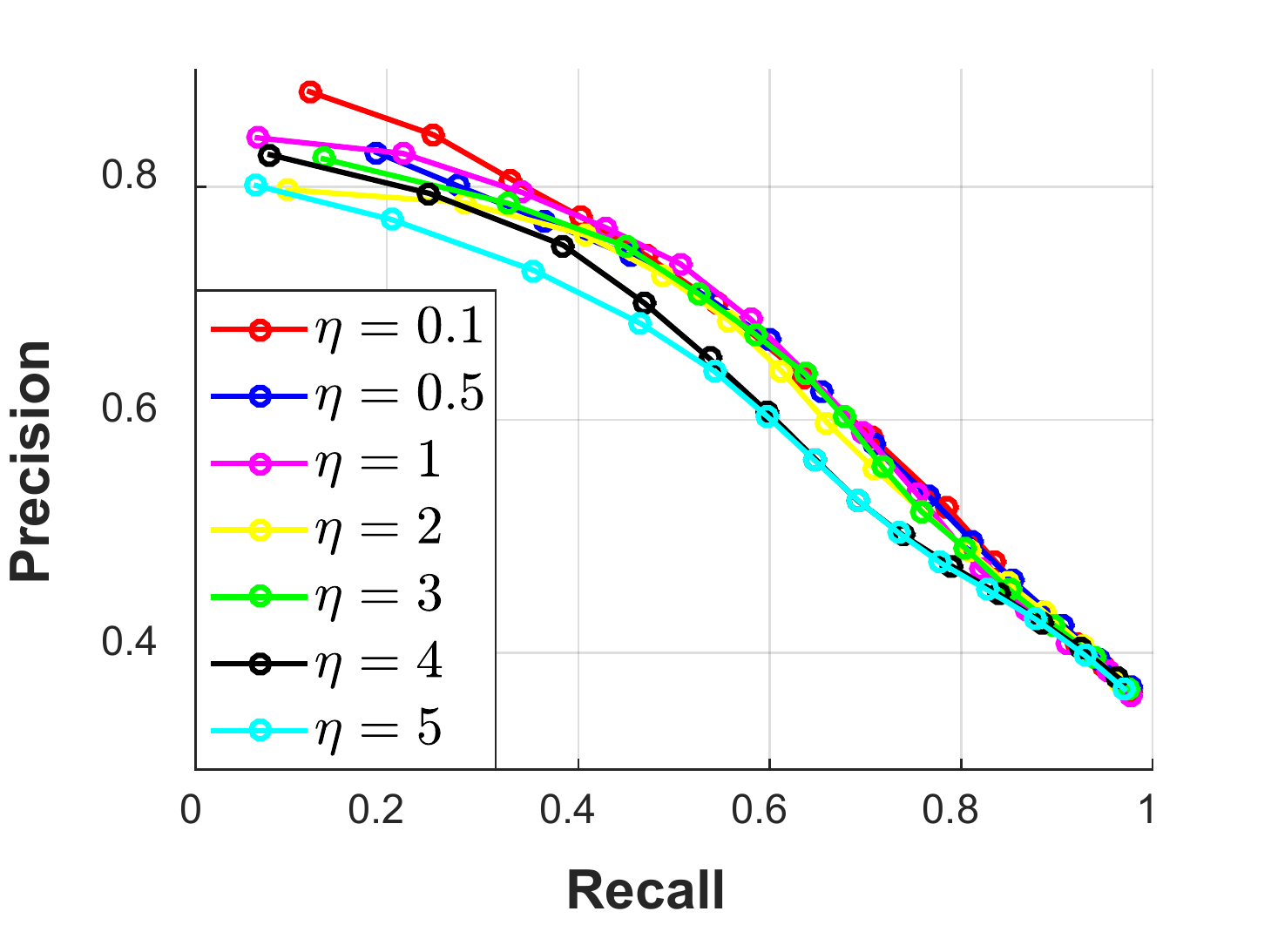}\\
    (d) Text $\to$ Image @NUS-WIDE
\end{minipage}\vspace*{-0pt}
\end{tabular}
\vspace{-0.2cm}
\caption{Hyper-parameter $\eta$}
\label{fig:HyperParameter}
\centering
\small
\begin{tabular}{c@{ }@{ }c@{ }@{ }c@{ }@{ }c}
\begin{minipage}{0.23\linewidth}\centering
    \includegraphics[width=1\textwidth]{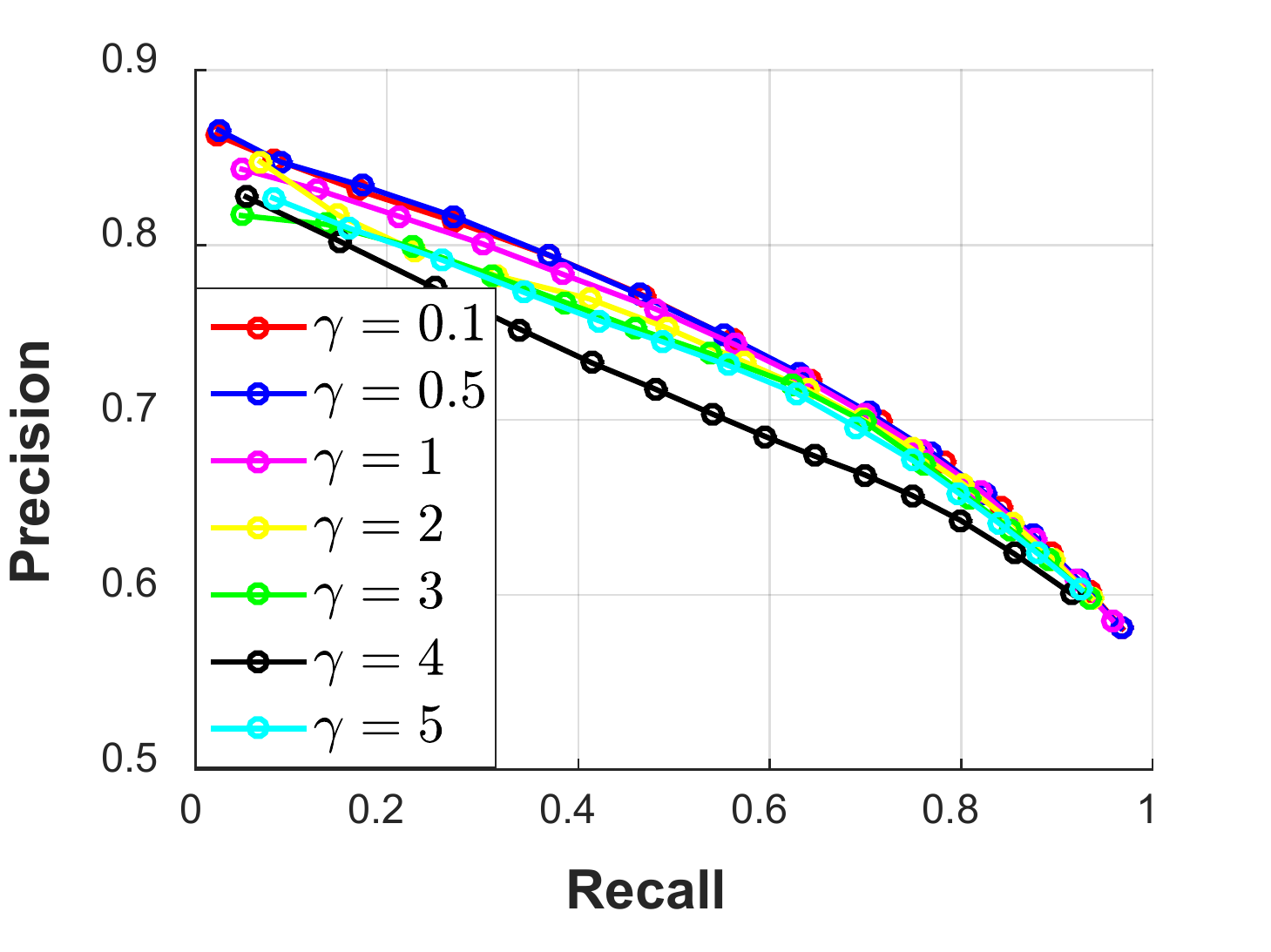}\\
    (a) Image $\to$ Text @MIRFLICKR-25K
\end{minipage} &
\begin{minipage}{0.23\linewidth}\centering
    \includegraphics[width=1\textwidth]{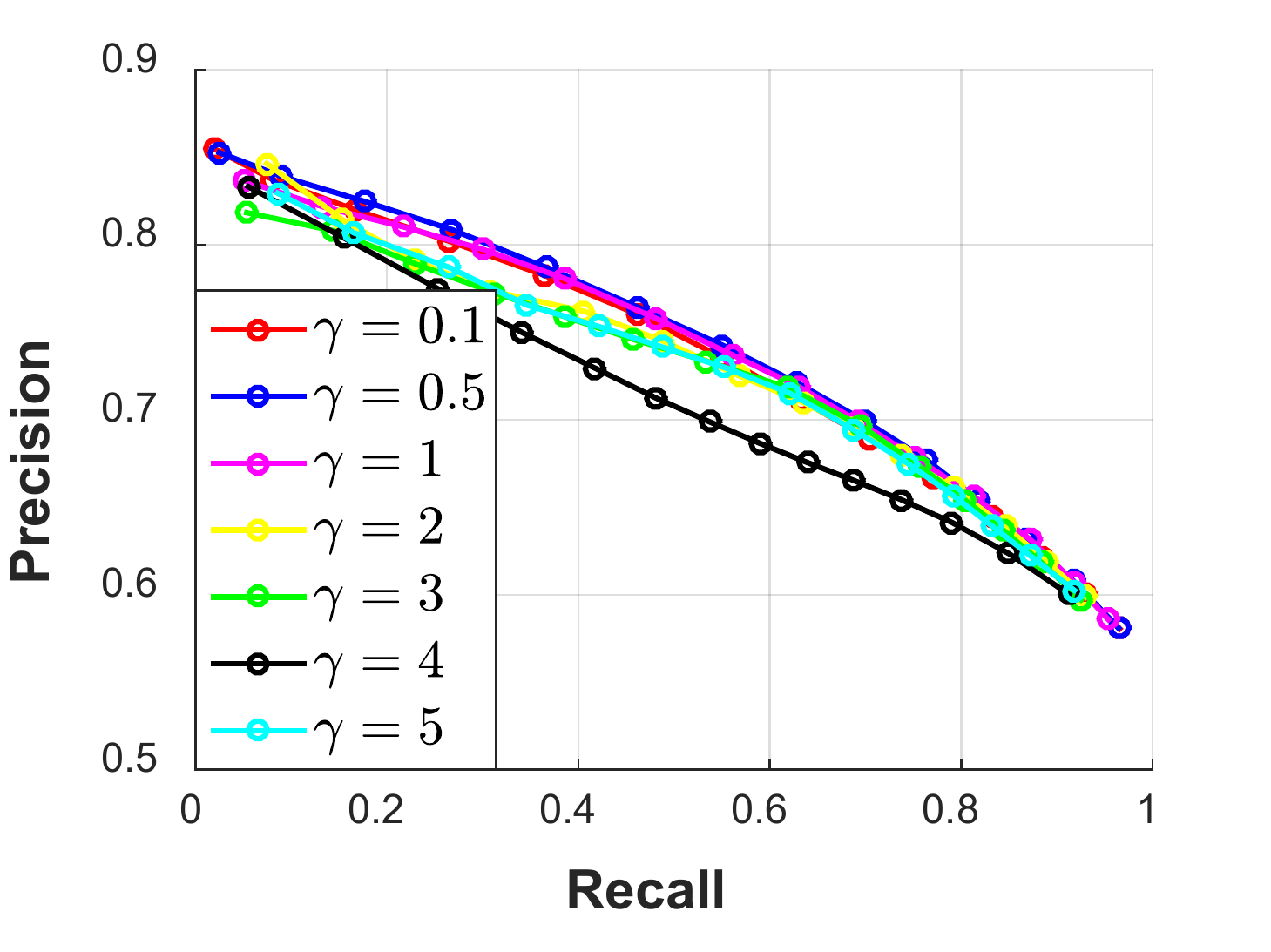}\\
    (b) Text $\to$ Image @MIRFLICKR-25K
\end{minipage} &
\begin{minipage}{0.23\linewidth}\centering
    \includegraphics[width=1\textwidth]{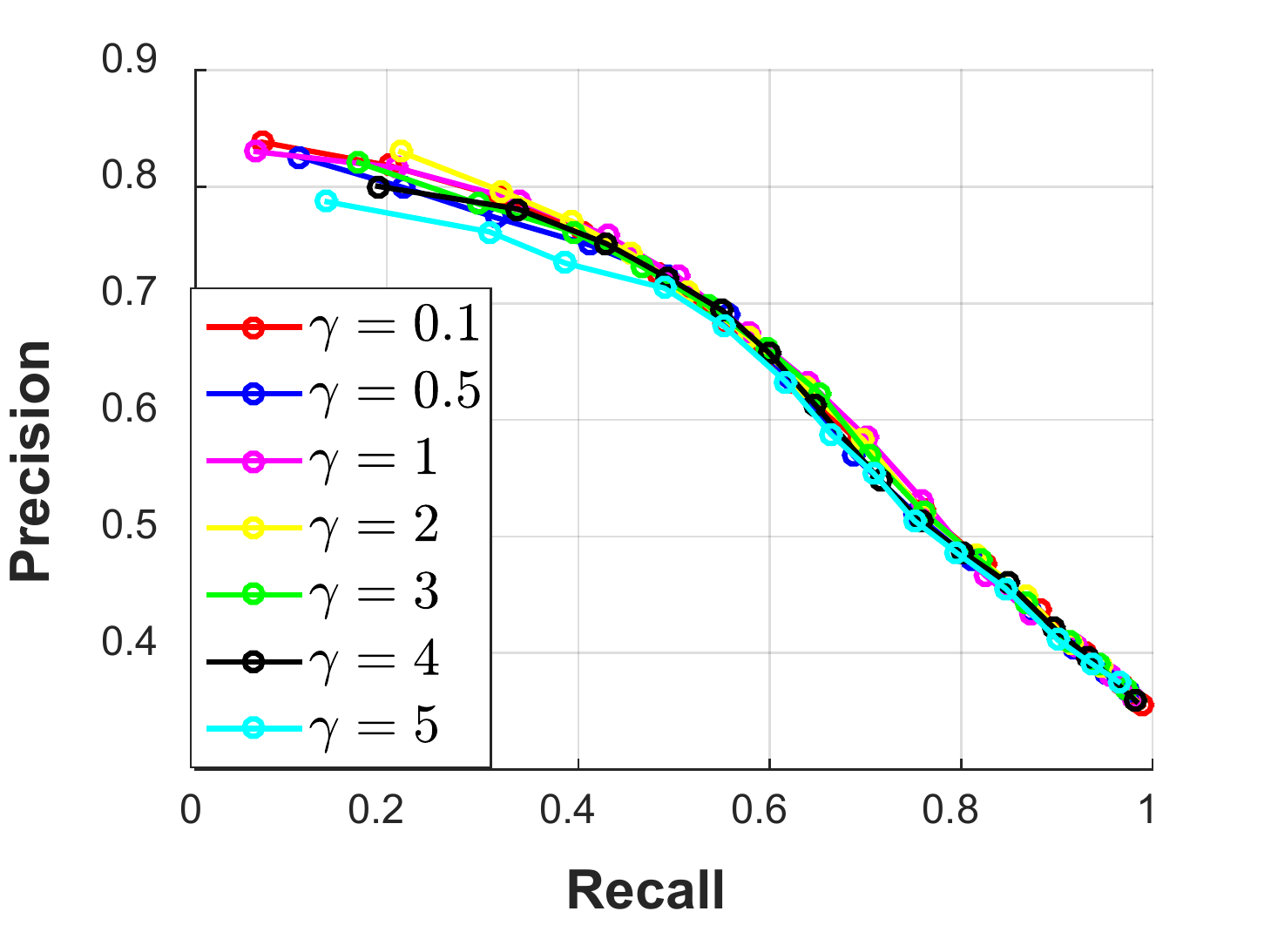}\\
    (c) Image $\to$ Text @NUS-WIDE
\end{minipage} &
\begin{minipage}{0.23\linewidth}\centering
    \includegraphics[width=1\textwidth]{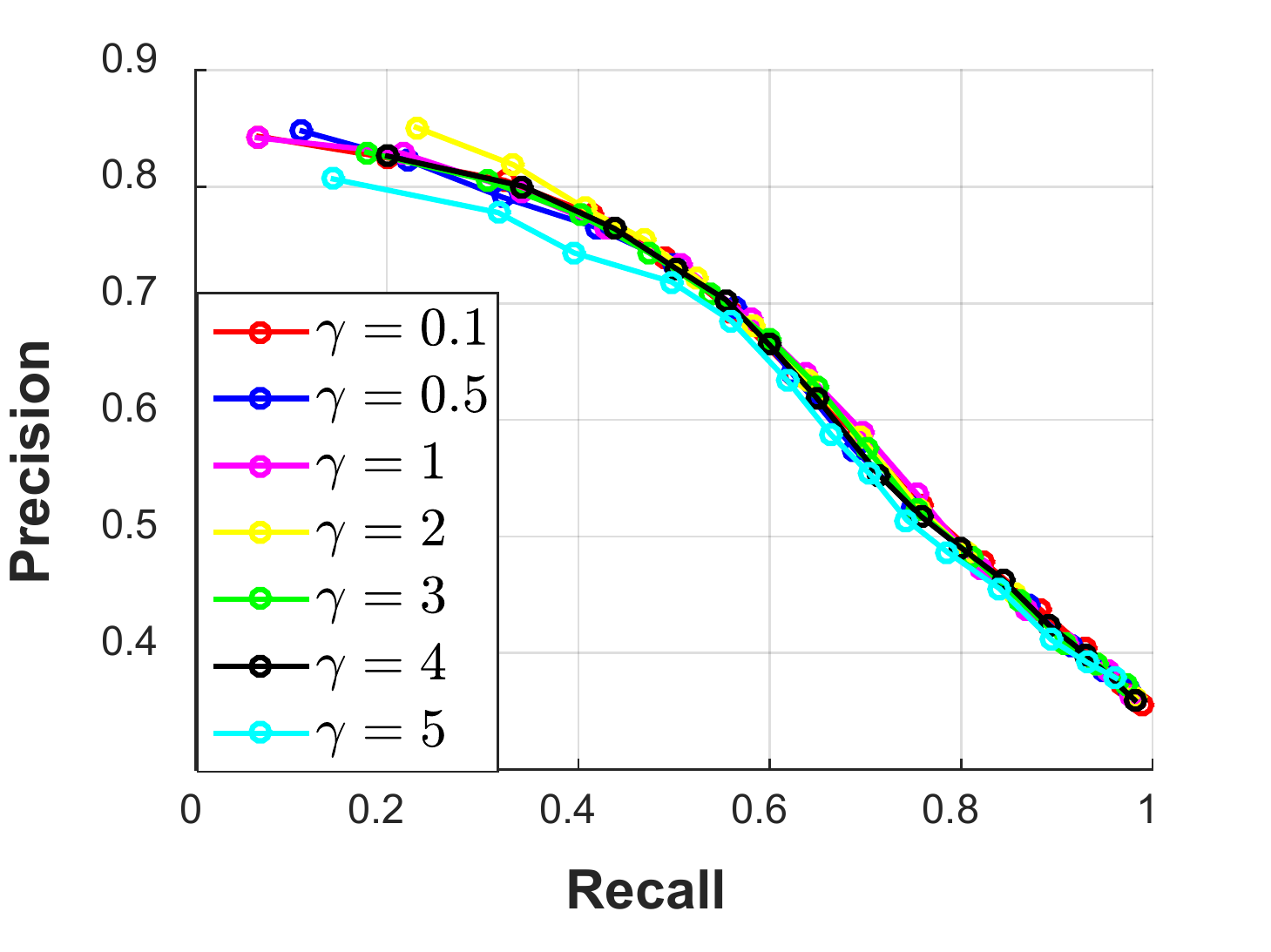}\\
    (d) Text $\to$ Image @NUS-WIDE
\end{minipage}\vspace*{-0pt}
\end{tabular}
\vspace{-0.2cm}
\caption{Hyper-parameter $\gamma$}
\label{fig:HyperParameter}
\end{figure*}

\end{document}